\documentstyle[twocolumn,prb,aps,epsf,floats]{revtex}

\begin{document}
\draft
\twocolumn[    
\hsize\textwidth\columnwidth\hsize\csname @twocolumnfalse\endcsname    
\title{Tuning a Josephson junction through a quantum critical point}

\author{ J.~K. Freericks$^{\dag}$, B.~K.~Nikoli\'c$^{\dag}$, and
P.~Miller$^*$ }

\address{
$^{\dag}$ Department of Physics, Georgetown University, 
 Washington, D.C. 20057-0995, U.S.A. \\
$^*$ Department of Physics, Brandeis University,
Waltham, MA 02454, U.S.A.}

\maketitle

\begin{abstract}
{
We tune the barrier of a Josephson junction through a zero-temperature
metal-insulator transition and study the thermodynamic behavior of the 
junction in the proximity of the quantum-critical point.  We examine a
short-coherence-length superconductor  and a barrier (that is described by a
Falicov-Kimball model) using the local approximation and 
dynamical mean-field theory. The inhomogeneous system is self-consistently
solved by 
performing a Fourier transformation in the planar momentum and exactly
inverting the remaining one-dimensional matrix with the renormalized
perturbation expansion.  Our results show a delicate interplay between
oscillations on the scale of the Fermi wavelength and pair-field
correlations on the scale of the coherence length, variations in the
current-phase relationship, and dramatic changes in the characteristic voltage
as a function of the barrier thickness or correlation strength (which can lead 
to an ``intrinsic'' pinhole effect). 
}
\end{abstract}

]

\section{Introduction}

The theoretical understanding of Josephson junctions has progressed 
dramatically over the past four decades\cite{Jose,likharev_review}.  
Recent advances~\cite{slreview} have been fostered by nanofabrication of 
superconducting mesoscopic devices,~\cite{katalskii} which, together with
high-temperature superconductor junctions,  have revived 
interest in the transport properties of superconductors weakly coupled
through a normal region. 
The interplay between phase-coherent electron propagation in the normal 
region and macroscopic phase coherence of Cooper pairs in
superconductors generates novel quantum interference phenomena since 
the proximity effect in such systems is mediated by a phase-coherent Andreev 
reflection~\cite{carlo_rmt,lambert}. However,
little attention has been paid to quantum effects on transport arising
from many-body correlations in the barrier separating the
superconductors. Such junctions are frequently encountered in high-$T_c$
systems where both superconducting electrodes and the normal region are
highly  correlated electronic systems.~\cite{htc} 

Low-$T_c$ junctions have 
large superconducting coherence lengths, and effects on the scale of the 
Fermi wavelength can usually be averaged over to accurately describe 
junctions by a quasiclassical (single-particle) approach. As the coherence
length of the superconductor becomes smaller and smaller (as in high-$T_c$
junctions) one can no longer ignore the interplay between oscillations
brought on by the Fermi surface and those due to pair-field correlations.
In addition, as junction sizes are made smaller and smaller, the barrier needs
to be tuned close to the metal-insulator transition in order to maintain
a large characteristic voltage (where properties of a Josephson junction
have been thought to be optimized\cite{Beas}).  The conventional 
proximity-effect theory cannot account for supercurrent transport in junctions
where the barrier approaches a metal-insulator transition\cite{htc}.
Therefore, these junctions must be described in a full many-body
approach that can properly account for the change in character of the
quantum mechanical system as the correlations drive a metal-insulator
transition.  
The standard single-particle approaches, like the full quantum
transport theories (scattering formalism~\cite{carlo_rmt,lambert,bardas},
and Green's-function techniques~\cite{levy1,levy2}) or
traditional quasiclassical Green's-function methods~\cite{schon}
are inadequate for this purpose (in general, the quasiclassical approaches
do not require a quasiparticle assumption\cite{ramer}, but the usual
quasiclassical Green's function, employed in nonuniform superconductivity
problems, can be expanded in terms of Andreev quasiparticle eigenfunctions and
energies\cite{yip}).

Recent progress in the dynamical mean-field theory\cite{DMFT1}
has shown how to generalize
the local approximation to inhomogeneous systems\cite{Pott} and to Josephson 
junctions\cite{miller}. Here we utilize this formalism to examine what happens 
as the barrier material is tuned through a quantum-critical transition
where the single-particle density of states is suppressed to zero and
a correlated metal-insulator transition occurs.  We find that in this region of
phase space, it is important to include self-consistency effects and
many-body effects.  
The simple analytical treatments~\cite{carlo_rmt} of Josephson junctions rely 
on the usage of rigid boundary conditions,~\cite{likharev_review} i.e., a step 
function model for the pair potential at a normal-superconductor interface. 
This is justified in narrow junctions (barrier width smaller than the bulk
coherence length $\xi_0$) where
the effect of the constriction induced by the narrow barrier on the order
parameter of the much wider superconductors is ``geometrically diluted'', or in 
wide junctions with high resistivity barriers (in both cases the critical 
current of the junction is much smaller than the bulk critical current of the
superconducting leads~\cite{sols_ferrer_1d}). On the other hand,
a self-consistent solution for the variation of the order parameter $\Delta(x)$
[i.e., pair-correlation function $F(x)=-\Delta(x)/U(x)$ with $U(x)$ the 
site-dependent interaction strength] induced by the current 
flow or geometry, not only ensures current conservation and allows one 
to find the 
critical current in an arbitrary geometry,~\cite{bagwell} but is unavoidable in 
situations where the proximity effect induces appreciable superconductivity 
in the normal region,~\cite{agrait} or when the thickness of the weak link is 
comparable~\cite{levy1} to $\xi_0$. Thus, the microscopic
self-consistent calculations\cite{levy1}
reveal a variation of $\Delta$ on length scales 
(like $\lambda_F$, the Fermi wavelength) smaller 
than $\xi_0$ (which is also of importance in 
high-$T_c$ junctions where the quasiclassical
approximation,~\cite{schon} $\xi_0 \gg \lambda_F$, does not hold). Our junctions
are wide, and even in the tunneling limit (i.e., with a correlated insulator 
barrier), they require a self-consistent
treatment because the many-body effects prevent a description in 
terms of simple phenomenological parameters (like the barrier transparency). 
Both self-consistency effects and many-body correlations are automatically 
included via the dynamical mean-field theory.  

Our results should shed light on high-$T_c$ superconductors even though
we are restricting ourselves to $s$-wave symmetry order parameters.  
This is because
the high-$T_c$ superconductors have short coherence lengths (on the order of 
a few lattice spacings) and have barrier materials [either from grain 
boundaries, ion-damage, or doping (such as Co-doping)] that are correlated
and lie close to the Mott metal-insulator transition\cite{htc}.  Our examination
of $s$-wave superconductors in this limit illuminates this new physical
regime without adding the complicated geometrical effects that arise from
$d$-wave order parameters (which will be investigated in a future study).

In Section II we briefly describe the formalism that is used in our
computational techniques.  Section III contains results on tuning through
the quantum critical point by increasing the correlation energy at a fixed
barrier thickness.  We examine four cases: (i) thin barrier; (ii) bilayer
barrier; (iii) barrier on the order of the bulk coherence length; and (iv)
thick barrier. In Section IV we tune the metal-insulator transition
by increasing the thickness of the barrier at fixed correlation energy.
We examine a weakly-correlated metal barrier, a strongly correlated
metal, and a Mott insulator, finding deviations from quasiclassical
results for the correlated insulator. Our conclusions are presented in 
Section V.

\section{Formalism}

\begin{figure}[htbp]
\epsfxsize=3.0in
\centerline{\epsffile{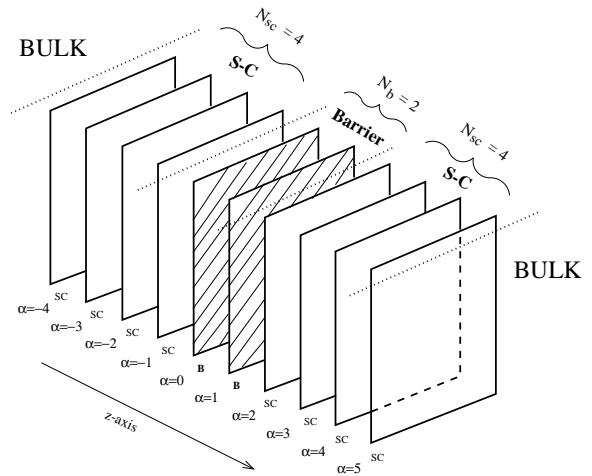}}
\caption{
\label{fig: planes}
Microscopic stacked planar geometry of a Josephson junction.
The sandwich of $N=10$ planes; $N_{SC}=4$ superconducting planes
coupled to a bulk superconductor on the left and $N_{b}=2$ barrier
planes on the right, followed by a further $N_{sc}=4$ superconducting planes
coupled to another bulk superconductor on the right.
The junction is allowed to have spatial inhomogeneity only within the $N$
modeled planes, but the calculations are for an infinite system.
In our calculations we always take $N_{SC}=30$ and $N_b$ ranges from 1
to 80.}
\end{figure}

The computations require a self-consistent calculation of the properties
of a Josephson junction within a many-body formalism.  To start, we
need to have a solution of the bulk superconductor,
which will provide the ``boundary condition'' for the simulations
in the bulk boundaries of the junction.  The bulk problem can be solved
directly in both the absence of a supercurrent and in the presence of
a supercurrent (where there is a uniform variation in the phase of
the superconducting order parameter).  The uniform bulk solution is then
employed to provide the boundary conditions for the junction beyond the
region where we determine properties self-consistently.  The inhomogeneous
problem consists of $N$ self-consistent planes embedded in the bulk
superconductor on each side (see Fig.~\ref{fig: planes}).  
The self consistent region
consists of a sandwich of $N_b$ barrier planes surrounded by $N_{SC}$
planes on each side $N=N_b+2N_{SC}$ (the word ``barrier'' is used since the
material through which the weak link between superconductors is made will
have its properties tuned from a metal to an
insulator).  In our solutions we choose $N_{SC}=30$
and $N_b$ ranges from 1 to 80.  Since the coherence length of the 
superconductor is $\xi_0=\hbar v_F/(\pi\Delta)\approx 4a$ (with $a$ the lattice
spacing) the self-consistent
superconducting region is approximately eight times the bulk
coherence length, which we find to be sufficient for our calculations.
This approach is useful, since it does not require us to make any assumptions 
about the boundary conditions at the interface between the barrier and
the superconductor, since they are determined self-consistently. The
approximation is the presence of a (typically small) discontinuity in the
supercurrent at the bulk superconductor--self-consistent-superconductor 
interface.  We have found that
the superconducting order has always healed to its bulk value at that point,
but sometimes there can be a jump of the superconducting phase
when one nears the critical current of the junction.
This discontinuity in the phase (corresponding to a breakdown of current
conservation at the bulk superconductor-simulated superconductor
interface) can become large for thick insulating barriers  or when one lies
on the decreasing current side of the current-phase diagram (see below).

We simulate an inhomogeneous system of stacked square lattice planes
that correspond to the superconductor-barrier-superconductor sandwich of
a Josephson junction. A lattice site corresponds to a unit cell (which
we normally picture as being one atomic site), and we assume a tight-binding
picture with the same hopping integral $t_{ij}$ between atomic sites within
a plane and atomic sites between the planes.  This description implies that
we are assuming the ``bare'' kinetic energy of the superconductor and the
barrier are identical (note that the renormalized density of states can be very
different, especially when the barrier is a correlated insulator).  Such a 
condition is not necessary in this formalism,
but we include it for simplicity, since it reduces the number of 
parameters that are varied
in the junction.  The superconductor is described by an attractive
Hubbard model\cite{hubbard}
in the Hartree-Fock approximation.  This is equivalent to 
the conventional Bardeen-Cooper-Schrieffer\cite{BCS} (or more accurately, the
Boguliubov-de Gennes\cite{DeGe} [which involves a nonconstant 
density of states due to the tight-binding approach])
description, except in this case the energy cutoff is determined by the 
electronic bandwidth rather than the phonon frequency.  In fact, the 
attractive Hubbard model offers richer behavior (that is not employed
in this contribution) showing a crossover to preformed pairs.  By including
higher-order processes in $U$, through a $T$-matrix\cite{nozieres} or
a dynamical mean-field theory\cite{keller} approach, one can study the
crossover\cite{legget} from BCS superconductivity, where pair formation and
condensation occur at $T_c$, to preformed pairs that condense at a lower
temperature (this should be important in short-coherence-length
superconductors like the high-$T_c$ materials).
The barrier
is described by a Falicov-Kimball model\cite{falicov_kimball}
at half filling.  This model
has two kinds of particles: (i) mobile electrons and (ii) static ions.
The average concentration of electrons is one per site and the average 
concentration of ions is one-half per site.  When an electron and an ion
occupy the same lattice site, there is a Coulomb attraction $U_{FK}$ between
them.  One can view this system as a binary alloy of $A$ and $B$ ions 
at 50\% concentration with $U_{FK}$ being the difference in site energy
between the $A$ and $B$ ionic sites (the off-diagonal energy is assumed to
be the same for the $A$ and $B$ ions).  The many-body problem is solved by 
taking an annealed average and is essentially the simplest disorder problem
(and the simplest many-body problem).
It undergoes a metal-insulator transition in the bulk\cite{vandongen}
(see below) which is why we adopt it for study here.

The Hamiltonian for the Josephson junction is then
\begin{eqnarray}
H&=&\sum_{ij\sigma}t_{ij}c_{i\sigma}^{\dag}c_{j\sigma}+\sum_iU_i\left (
c_{i\uparrow}^{\dag}c_{i\uparrow}-\frac{1}{2}\right ) \left (
c_{i\downarrow}^{\dag}c_{i\downarrow}-\frac{1}{2}\right )\cr
&+&\sum_{i\sigma}U_i^{FK}c_{i\sigma}^{\dag}c_{i\sigma}\left ( w_i-\frac{1}{2}
\right ),
\label{eq: hamiltonian}
\end{eqnarray}
where $c_{i\sigma}^{\dag}$ ($c_{i\sigma}$) creates (destroys) an electron
of spin $\sigma$ at site $i$, $t_{ij}$ is the hopping integral between 
nearest neighbor sites $i$ and $j$ (we measure energies in units of $t$),
$U_i=-2$ is the attractive Hubbard interaction for sites within the
superconducting planes, $U_i^{FK}$ is the Falicov-Kimball interaction
for planes within the barrier, and $w_i$ is a classical variable that
equals 1 if an $A$ ion occupies site $i$ and is zero if a $B$ ion
occupies site $i$.  A
chemical potential $\mu$ is employed to determine the filling.  Since
we work at half filling for both the superconductor and the barrier, we have
$\mu=0$. Note that if $U_i=U_i^{FK}=0$ for all lattice sites, the Hamiltonian
describes tight-binding electrons on a simple cubic lattice.

\begin{figure}[htbp]
\epsfxsize=3.0in
\centerline{\epsffile{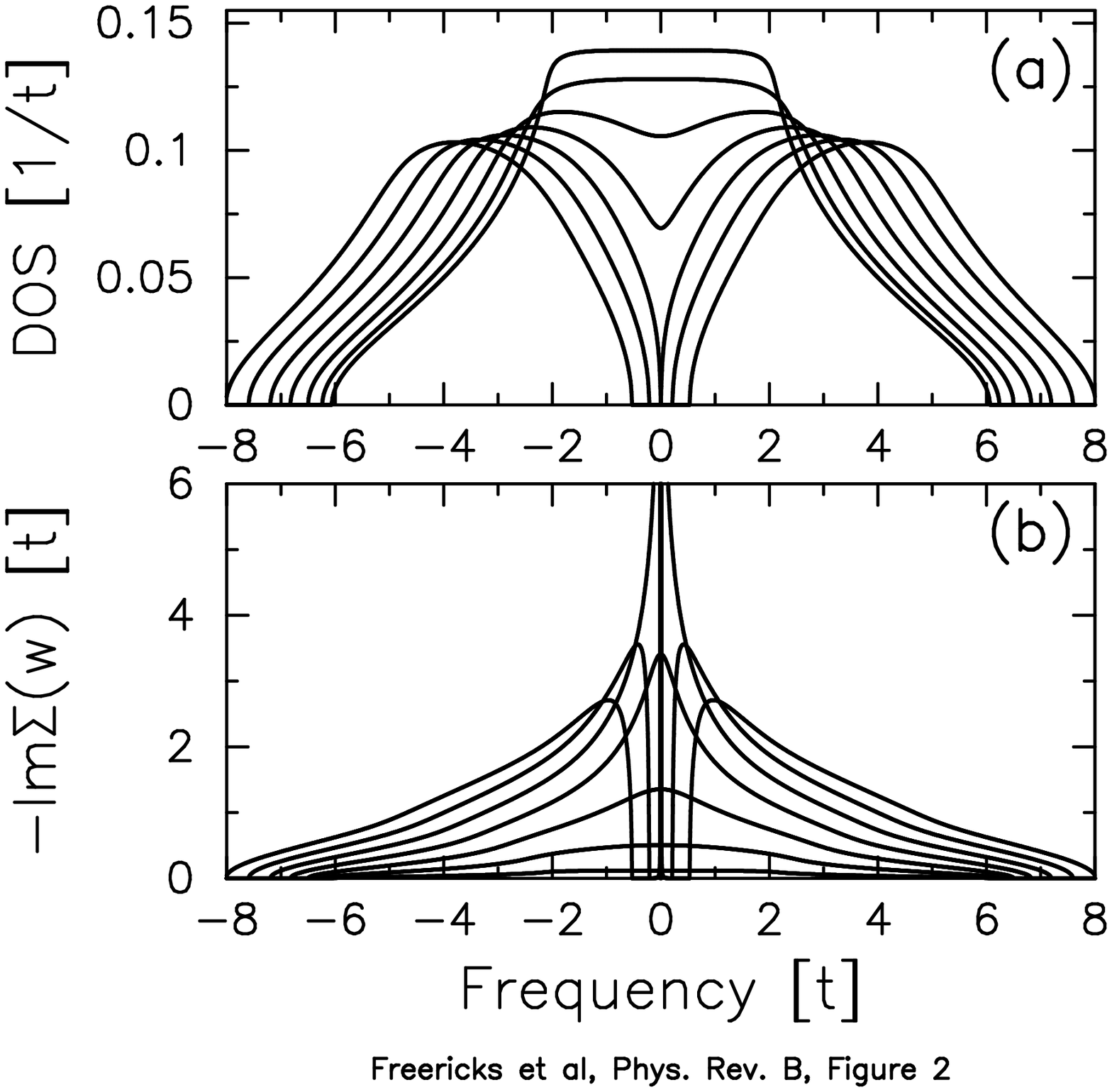}}
\caption{
\label{fig: fkbulk}
(a) Electronic density of states for the bulk barrier (simple-cubic
lattice) described by the Falicov-Kimball model in the local approximation.  
The value of $U_{FK}$ ranges from 1 to 7 in steps of 1.  As $U_{FK}$
increases the density of states first
develops a pseudogap and then a real gap.
(b) Absolute value of the 
imaginary part of the local retarded self energy for low frequency on
the real axis.  See how the
curvature has the wrong sign for a Fermi liquid and how the imaginary
part diverges at zero frequency
as one goes through the quantum-critical point and a pole
develops in the self energy. }
\end{figure}

The superconducting regions are described by an attractive Hubbard model
with $U_i=-2$ and $w_i=0$ for all superconductor sites. The homogeneous
bulk superconductor has a transition temperature $T_c=0.11$ and a 
zero-temperature order parameter $\Delta=0.198$. This yields a standard
BCS gap ratio $2\Delta/(k_BT_c)\approx 3.6$ and a coherence length
$\xi_0=\hbar v_F/(\pi\Delta)$ that ranges from $3.5a$ to $4.3a$ depending
on whether we average the absolute value of $v_F$ over the Fermi surface
or take the root-mean-square of $v_F$ (a cubic lattice at half-filling has
a direction-dependent Fermi velocity); a fit of the decay of the superconducting
order as it is disturbed at the superconductor-barrier interface\cite{miller}
gives $\xi_0\approx 3.7a$. The bulk critical current per unit area is 
$I_{c,{\rm bulk}}=0.0289 (2et)/(\hbar a^2) $.  The value of our bulk critical 
current density is slightly higher than the one determined by
a Landau depairing velocity $v_d=\Delta/\hbar k_F$ 
($j_{c,{\rm bulk}}=env_d$, where the density of particles is $n=k_F^3/2\pi^2$, 
assuming a spherical Fermi surface) because of the possibility to have 
gapless superconductivity in three dimensions at superfluid velocities slightly
exceeding~\cite{bardeen} $v_d$ (note that $k_F$ is direction-dependent for a 
cubic lattice at half-filling). Calculations on our junction are performed at
a temperature of $T=0.01$, which is effectively at the zero-temperature
limit ($T/T_c\approx 0.09$) for the superconducting properties.  The barrier 
region is described by a half-filled
Falicov-Kimball model in the symmetric limit of $\langle w_i \rangle=0.5$.
In the bulk, this barrier undergoes a metal-insulator transition at
$U_{FK}\approx 4.9$ (since the bandwidth of the simple cubic lattice is 12,
the metal-insulator transition occurs when $U_{FK}$ is on the order of one-half
of the bandwidth).  This is illustrated in Fig.~\ref{fig: fkbulk}(a) where we
show the single-particle density of states for a bulk barrier as a function
of $U_{FK}$.  The density of states for this model is independent of
temperature\cite{vandongen}.  Since the system is not a Fermi liquid for 
nonzero $U_{FK}$,
one can see the density of states first develops a pseudogap and then
is suppressed entirely to zero as the correlations are increased and it
becomes a correlated insulator.  The opening of the gap is continuous.
In Fig.~\ref{fig: fkbulk}(b) we show the imaginary part of the local self
energy at low energies.  This result is also temperature 
independent\cite{vandongen}.  We see
that the curvature of the self energy has the wrong sign in the metallic
regime (which is one reason why it is not a Fermi liquid) and that it
diverges (and becomes a delta function) as the system crosses over into
the insulating phase.  This occurs because the self energy develops a
pole at zero energy in the insulating phase.  Such behavior can only be 
seen in a many-body treatment of the system.

The inhomogeneous system is solved by employing the matrix formulation
of Nambu\cite{Namb} for the Green's function $G({\bf r}_i,{\bf r}_j,i\omega_n)$
for two lattice sites ${\bf r}_i$ and ${\bf r}_j$ at the Matsubara
frequency $i\omega_n=i\pi T(2n+1)$,
\begin{equation}
G({\bf r}_{i},{\bf r}_{j},i\omega_{n}) =
\left( \begin{array}{cc}
G({\bf r}_{i},{\bf r}_{j},i\omega_{n}) &
F({\bf r}_{i},{\bf r}_{j},i\omega_{n}) \\*
\overline{F}({\bf r}_{i},{\bf r}_{j},i\omega_{n}) &
- G^{*}({\bf r}_{i},{\bf r}_{j},i\omega_{n})
\end{array} \right) ,
\label{eq: g_nambu}
\end{equation} 
and the corresponding local self energy,
\begin{equation}
\Sigma ({\bf r}_{i},i\omega_{n}) =
\left( \begin{array}{cc}
\Sigma({\bf r}_{i},i\omega_{n}) & \phi({\bf r}_{i},i\omega_{n}) \\*
\phi^{*}({\bf r}_{i},i\omega_{n}) & - \Sigma^{*}({\bf r}_{i},i\omega_{n})
\end{array} \right).
\label{eq: sigma_nambu}
\end{equation}   
The diagonal and off-diagonal Green's functions are
defined respectively as:
\begin{eqnarray}
G({\bf r}_{i},{\bf r}_{j},i\omega_{n}) & = &
-  \int^{\beta}_{0}
d\tau
\exp (i\omega_{n}\tau)
\langle{\rm T}_{\tau}
\hat{c}_{j\sigma}(\tau) \hat{c}^{\dagger}_{i\sigma}(0) \rangle,
\\*
F({\bf r}_{i},{\bf r}_{j},i\omega_{n}) &
= & - \int^{\beta}_{0} d\tau
\exp (i\omega_{n}\tau)
\left\langle{\rm T}_{\tau}
\hat{c}_{j\uparrow}(\tau) \hat{c}_{i\downarrow}(0) \right\rangle,
\label{eq: g_f_def}
\end{eqnarray}
where ${\rm T}_{\tau}$ denotes time-ordering in $\tau$ and
$\beta=1/T$.

The self energies and Green's functions are coupled together through
Dyson's equation,
\begin{eqnarray} 
\lefteqn{G({\bf r}_{i},{\bf r}_{j},i\omega_{n})= G^0({\bf
r}_{i},{\bf r}_{j},i\omega_{n})} \nonumber \\
& & \mbox{} +
\sum_{l} G^0({\bf r}_{i},{\bf r}_{l},i\omega_{n}) \Sigma({\bf
r}_{l},i\omega_{n}) G({\bf r}_{l},{\bf r}_{j},i\omega_{n}),
\label{eq: Dys}
\end{eqnarray}
where we have included the local approximation for the self energy,
$\Sigma({\bf r}_{i},{\bf r}_{j},i\omega_{n})
=  \Sigma({\bf r}_{i},i\omega_{n})\delta_{ij}$. 
The non-interacting Green's function,
$G^0({\bf r}_{i},{\bf r}_{j},i\omega_{n})$
is diagonal in Nambu space, with upper diagonal component given by:
\begin{equation}
G^{0}({\bf r}_{i},{\bf r}_{j},i\omega_{n}) = \int d^{3}{\bf k}
\frac{ \mbox{\rm e}^{i {\bf k}\cdot({\bf r}_{i}-{\bf r}_{j}) } }
{ i\omega_{n} +\mu-\varepsilon_{{\bf k}} }.
\end{equation}
We emphasize that $G^0$ is the non-interacting
Green's function and is {\it not} the effective medium of an
equivalent atomic problem.           

Details of the computational scheme have been described 
elsewhere\cite{miller}.
Here we simply summarize the algorithm.  The junction is inhomogeneous
in the $z$-direction only, since it has translational symmetry within
each plane.  The algorithm begins by converting the three-dimensional
system to a quasi-one-dimensional system using the method of
Potthoff and Nolting\cite{Pott}.  We perform a Fourier transformation within 
each plane to determine the mixed-basis Green's function [defined in terms
of two-dimensional momenta ($k_x$ and $k_y$) and the $z$-coordinate of
the plane] under the assumption that the electronic self energy is
local (but can vary from plane to plane). For each momentum in the 
two-dimensional Brillouin zone, we
have a one-dimensional problem with a sparse matrix, since the only coupling
between planes is due to the hopping to each neighboring plane.  The
infinite ``tridiagonal'' matrix can be inverted by employing the 
renormalized perturbation expansion\cite{Econ}, which calculates both the 
single plane and the nearest neighbor Green's functions. A final summation 
over the two-dimensional momenta produces the local Green's function and the
Green's function for propagation from one plane to its neighboring plane.
The dynamical mean-field theory is then employed to calculate the local
self energy from the local Green's function and then the local Green's function
is calculated from inverting the quasi-one-dimensional matrix.  These two
steps are repeated until the Green's functions have converged to a fixed
point. At the fixed point, we have a self-consistent solution of the
inhomogeneous problem that allows for nonuniform variations in both
the pair-field correlations (or equivalently the superconducting order
parameter) and in the phase.  One important consistency check 
is total current conservation at each plane in the self-consistent 
region.  All calculations conserve current except in extreme
cases for thick insulating barriers (see below). But there can be
discontinuities in the current at the bulk-superconductor--self-consistent
superconductor interface (since this is far from the Josephson junction,
it has a negligible effect on the results).
This computational algorithm is a generalization of the conventional
Boguliubov-de Gennes approach to allow for correlations within the
barrier.

This algorithm can be performed for the normal state or for the superconducting
state and can be performed on the imaginary or real frequency axes. We
work on the real axis in order to calculate the normal state
resistance.  Since we have a many-body system, we must use Kubo's
formula for the conductivity.  Details for this calculation appeared
elsewhere\cite{miller}.  Our formalism calculates the conductivity by neglecting
vertex corrections and evaluating the simple bubble diagram (which becomes
exact in the infinite-dimensional limit\cite{khurana}).

\section{Tuning the correlation strength through a metal-insulator transition}
\label{sec:tuning_ufk}

We begin by presenting results for a fixed barrier thickness, and vary the
Falicov-Kimball coupling strength.  We study four different systems:
(i) a thin barrier with $N_b=1$; (ii) a bilayer barrier with $N_b=2$; (iii)
a barrier on the order of the bulk superconducting
coherence length $N_b=5$; and (iv) a thick barrier $N_b=20$.  

\subsection{Thin barrier $(N_b=1)$}

A single plane barrier must be in the very strong insulating limit before
it can severely affect the transport perpendicular to the plane.  Hence
we expect to have to increase the Falicov-Kimball interaction to be much larger
than 5 before the junction starts to display ``insulating'' behavior.  
Similarly, we expect the critical current to be close to the bulk critical
current, because the plane is so thin (at least for metallic barriers).  In 
this regime, self-consistency
is critical in determining the properties of the junction\cite{sols_ferrer_1d}. 

\begin{figure}[htbp]
\epsfxsize=3.0in
\centerline{\epsffile{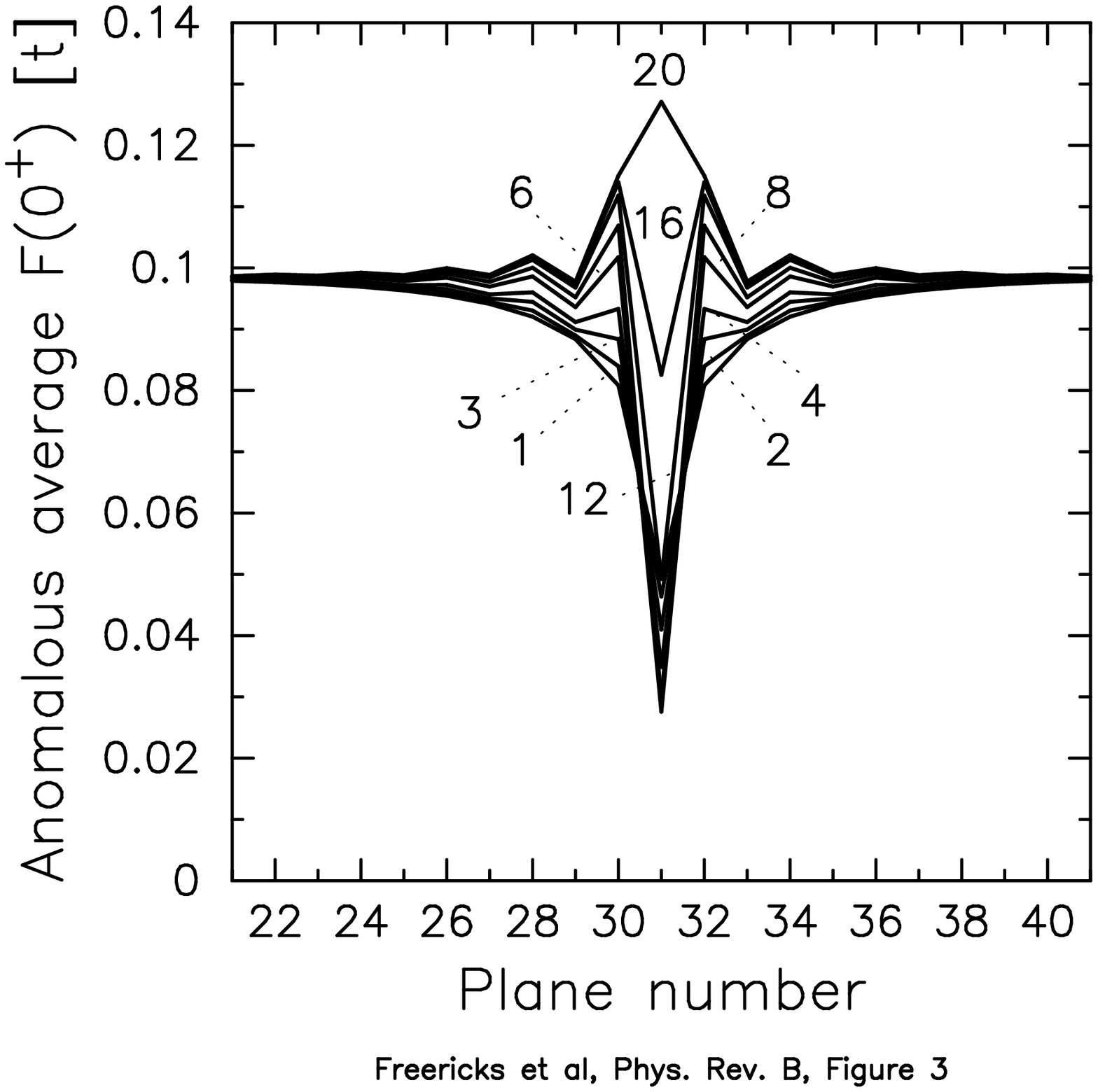}}
\caption{
\label{fig: n=1_f}
Proximity effect for a thin barrier ($N_b=1)$.  The anomalous average is
plotted versus plane number (the insulating barrier lies at plane 31).
The numbers indicate the value of $U_{FK}$ (1,2,3,4,6,8,12,16, and 20);
the anomalous average monotonically increases with $U_{FK}$ in the range
between planes 29 and 30.
Note how oscillations develop as the correlations increase until the
superconductivity ultimately becomes {\it enhanced} in the barrier for the
strong insulator.}
\end{figure}

We begin by examining the proximity effect within the junction.  Since the
Hubbard attraction is zero within the barrier, the superconducting gap
$\Delta$, which is proportional to the Hubbard attraction,
identically vanishes there.  But we can still examine the
superconducting pair-field correlations by plotting the anomalous average
at equal times $F(\tau=0^+)$.  This Green's function is continuous as one
passes through the superconductor-barrier interface.  We show $F(0^+)$ in
Fig.~\ref{fig: n=1_f}.  Notice how the correlated metal ($U_{FK}<2.5$) appears 
just 
as we expect it to: the superconductivity is smoothly depressed as we approach
the barrier and then decreases within the barrier as correlations increase.
As the bulk barrier enters the pseudogap regime ($2.5<U_{FK}<5$) we see small
oscillations appear in the superconductor, and the superconductivity continues
to be depressed within the barrier.  In the correlated insulator regime
($5<U_{FK}<8$) the oscillations continue to grow and the superconductivity 
within
the barrier is small, but rather insensitive to $U_{FK}$.  In the strong 
insulating
regime ($U_{FK}>8$), we find that the oscillations become large and the 
superconductivity eventually becomes {\it enhanced within the barrier}!  We 
believe that the oscillations and this 
enhancement of the anomalous average
are arising from a surface effect of the superconducting 
half-planes---each half-plane develops oscillations near
the surface (as the barrier becomes more insulating). 

\begin{figure}[htbp]
\epsfxsize=3.0in
\centerline{\epsffile{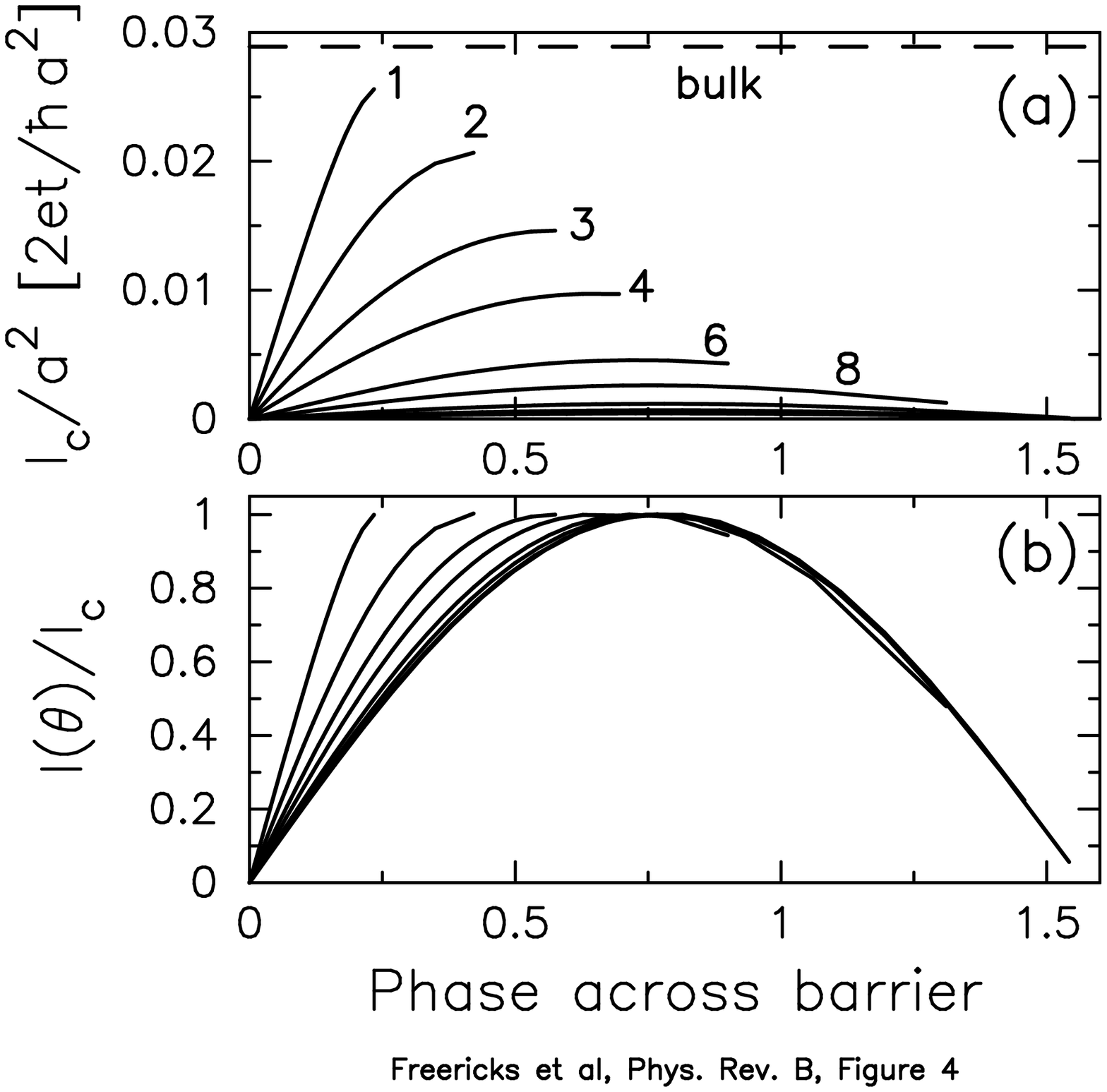}}
\caption{
\label{fig: n=1_current}
Current-phase relation for a single-plane barrier ($N_b=1$).  (a) $I(\theta)$
plotted for various values of $U_{FK}$ (1,2,3,4,6,8,12,16, and 20); the numbers
denote the value of $U_{FK}$.  The bulk critical current is shown
for comparison as the horizontal dashed line.  Note how in the weakly correlated
limit the curve is nonsinusoidal, but becomes sinusoidal as the
correlations increase. (b) Normalized $I(\theta)/I_c$.  Note how this 
approaches $\sin 2\theta$ for strong correlation, as discussed in the text.
[The same values of $U_{FK}$ are plotted in (b), but we don't label the
curves here.]}
\end{figure}

In Fig.~\ref{fig: n=1_current}, we show plots of (a) the current-phase relation
as well as (b) the normalized relation $I(\theta)/I_c$. The phase difference
across the junction is defined as the total phase across the barrier plane.
Since there is only one plane, and the majority of the phase jump occurs
at the barrier plane, we find that the $N_b=1$ Josephson junction has 
significant phase change over the superconducting region, since we must 
define the barrier to
begin at a distance halfway between the last superconducting plane (on the left)
and the barrier plane and end halfway between the barrier plane and the 
first superconducting plane (on the right). This result arises from the fact 
that the barrier plane is so thin and because we have discretized real
space to correspond to the atomic unit cells.  Note how the shape of the 
current-phase relation is far from sinusoidal when the critical current
is close to the bulk critical current.  As the barrier becomes more
insulating, the critical current decreases and approaches a more sinusoidal 
shape.  Because of the lattice nature of the model,  
the current-phase relation actually approaches $I(\theta)=I_c\sin(2\theta)$
for a single-plane junction (note that this is an artifact of the 
coarse-graining in real space that maps the junction onto a lattice).  More 
metallic barriers have the maximum of
the current occur at a phase difference much smaller than $\pi/4$ as
predicted to occur in self-consistent calculations\cite{sols_ferrer_1d}, and 
seen in our previous work\cite{miller}.

\begin{figure}[htbp]
\epsfxsize=3.0in
\centerline{\epsffile{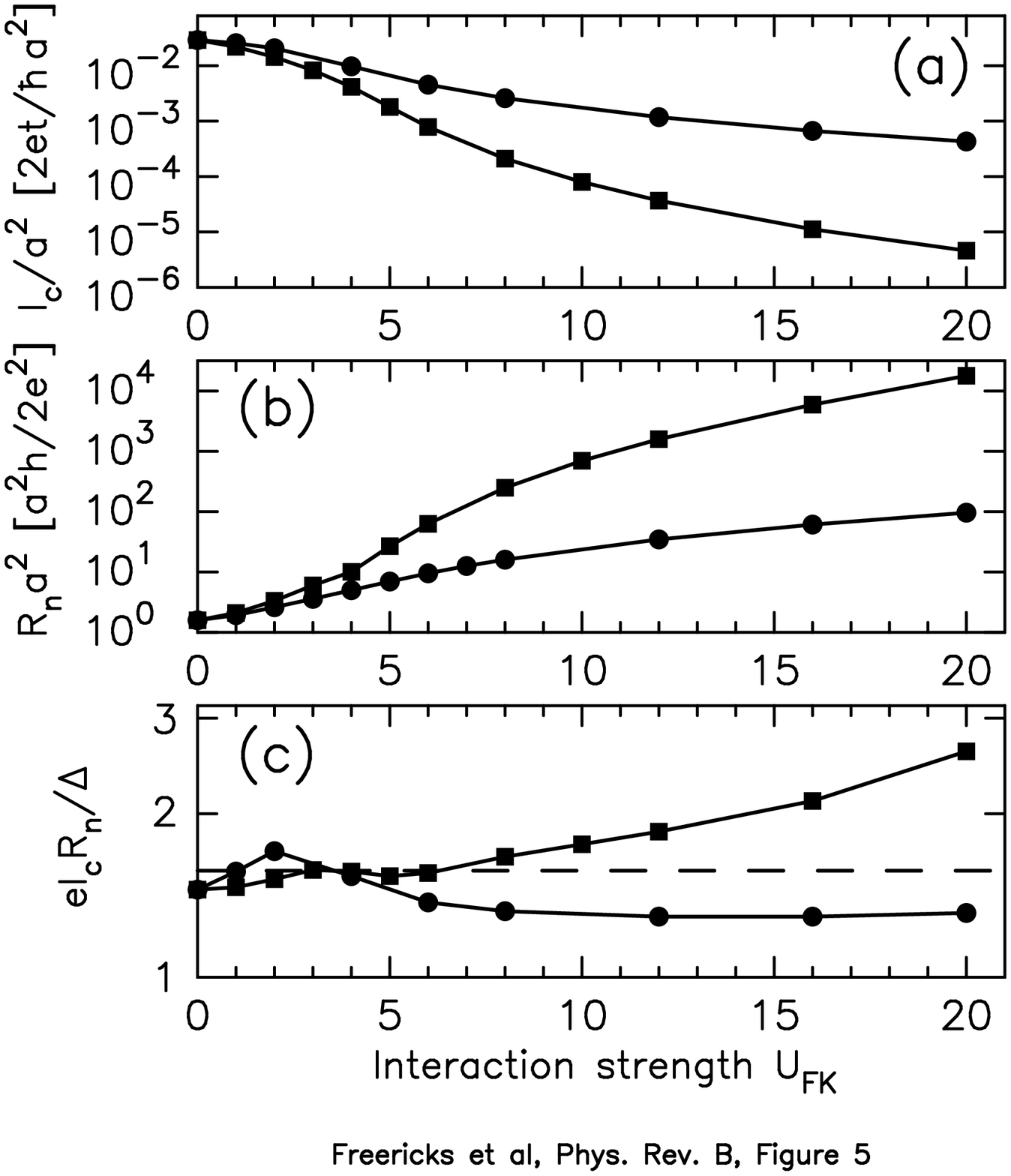}}
\caption{
\label{fig: n=1_2_icrn}
(a) Critical current, (b) normal state resistance, and (c) characteristic
voltage of the Josephson junctions as a function of the Falicov-Kimball
interaction within the barrier.  The circular symbols are for $N_b=1$ and the 
squares are for
$N_b=2$. Note how the critical current decreases and the junction
resistance increases as expected, and how the characteristic voltage does not
depend too strongly on the correlation strength. The dependence on correlation
strength for the
bilayer junction is much stronger than for the thin junction. The dashed line
in (c) is the Ambegaokar-Baratoff prediction.}
\end{figure}

Finally we show a plot of the Josephson critical current, the normal state
resistance $R_n$, and the characteristic voltage $I_cR_n$ for the single plane
barrier. The plots are on a semilogarithmic scale. The critical current drops
by about two orders of magnitude as $U_{FK}$ increases from 0 to 20.  This
occurs even though the anomalous average increases in the strongly
insulating limit! The plot of $R_n$ shows the expected increase as the
correlations increase.  But even at a large value of $U_{FK}=20$, the resistance
only increases by about two orders of magnitude over the noninteracting 
limit.  Finally, we show the characteristic voltage $I_cR_n$ in 
Fig.~\ref{fig: n=1_2_icrn}(c).  Its value
does not change much, but shows a mild optimization for the moderately
correlated metallic phase.  In the metallic limit $U\rightarrow 0$, the 
$I_cR_n$ product approaches the product of the bulk critical current
times the Sharvin resistance, which is $0.287t/e=1.45\Delta /e$.  This
result is different from the clean Kulik-Omelyanchuk limit\cite{kulik} 
of $\pi\Delta /e$ because we are treating a different geometry from a 
point contact (which can be described as a ``plane'' contact).  As
the correlations increase within the metallic phase, the characteristic 
voltage peaks for $U_{FK} \approx 2$ at a value somewhat smaller than the 
dirty limit
of the Kulik-Omelyanchuk formula\cite{kulik} for a superconducting point contact 
$I_cR_n=0.66\pi\Delta /e$ at $T=0$. 
In fact, there are two possibilities for the $I_cR_n$ product (i.e., critical 
current) of a short contact with diffusive scattering. Namely, in the 
single-particle picture scattering properties of a normal region can 
be described by the universal distribution 
of transparencies $D$ (defined as the distribution of
eigenvalues of the matrix ${\bf t}{\bf t}^\dag$, 
where ${\bf t}$ is the transmission matrix connecting incoming to outgoing 
transverse propagating modes\cite{carlo_rmt}) 
given by either the Dorokhov expression~\cite{dorokhov} 
$P_{\rm Do}(D)=(G/2G_Q) \, [D\sqrt{1-D}]^{-1}$ (valid for most
bulk conductors), or the
Schep-Bauer distribution\cite{schep-bauer}
$P_{\rm SB}(D)=(G/\pi G_Q) \, [D^{3/2}\sqrt{1-D}]^{-1}$ (valid for
sub-nm-thick barriers\cite{yehuda}).  Here, $G_Q=2e^2/h$ 
is the conductance quantum and $G=\int_0^1 dD\, P(D) D$ is the disorder-averaged
conductance. The total current is found by integrating the current carried by
a single channel $I(D)$ (with a transparency of $D$) over the distribution
function $P(D)$ as shown in  the multiple Andreev reflection 
theory\cite{bardas}. This integral, $I=\int_0^1 dD \, P(D)I(D)$,
then leads to the following characteristic voltages:
$I_cR_n=0.66 \pi \Delta/e$ for $P_{\rm Do}(D)$ and 
$I_c R_n=0.61 \pi \Delta/e$ for $P_{\rm SB}(D)$. 
We find that in the case of a single-plane barrier made of an FK correlated 
metal, the largest $I_cR_n$ (obtained for $U_{FK}=2$) is slightly below the 
value determined by $P_{\rm SB}(D)$. This can be attributed to effects of 
self-consistency (which always lower the critical current because of the 
depression of the order parameter in the superconducting leads due to the 
proximity effect), or to the fact that such an interface cannot be 
described by the Schep-Bauer distribution $P_{\rm SB}$ 
(the rationale behind the comparison of our barriers, dominated by many-body 
correlations, with conventional results relying on the single-particle picture 
of transport, is elaborated further in Sec.~\ref{sec:tuning_thick}).

As the junction barrier becomes more insulating, the
characteristic voltage becomes essentially constant as expected from the
Ambegaokar-Baratoff limit\cite{Ambe} $\pi\Delta /(2e)$ (dashed
line). But the magnitude of the characteristic
voltage is approximately 15\% smaller than that predicted by them
[$\pi\Delta/(2e)$ versus our calculated result of $1.31\Delta/e$].
Once again, this small reduction arises from the self-consistency for a
short-coherence-length superconductor which reduces the gap as one
approaches the barrier and from the Fermi-surface averaging of the
transport, since the Fermi surface is far from spherical at half-filling.

\subsection{Bilayer $(N_b=2)$}

\begin{figure}[htbp]
\epsfxsize=3.0in
\centerline{\epsffile{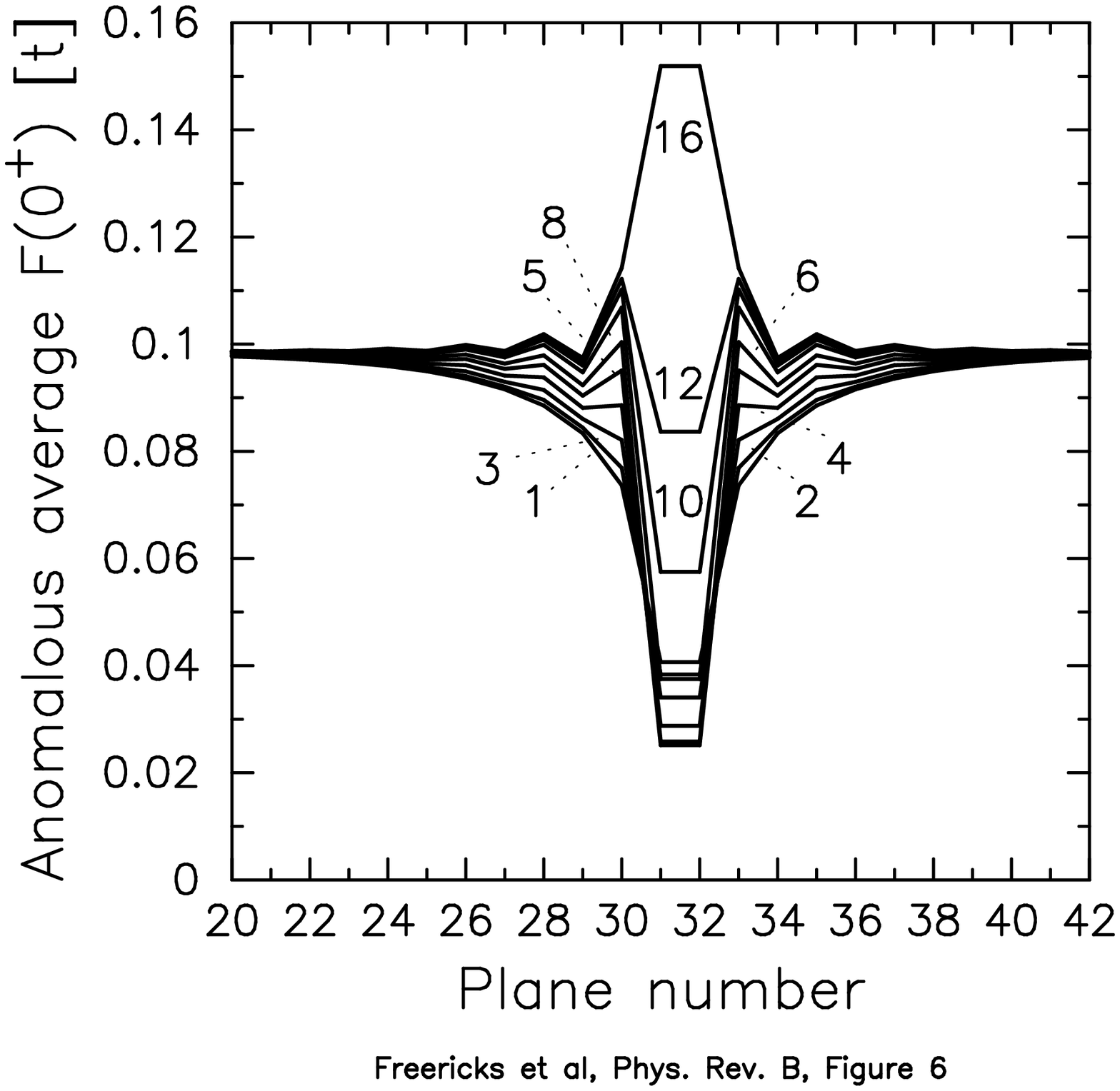}}
\caption{
\label{fig: n=2_f}
Anomalous average for differing correlation strengths and $N_b=2$.
Note how similar these results are to the thin barrier case---how the
anomalous average initially drops within the barrier, then oscillations 
develop, followed by an enhancement for the strongest correlation strengths.}
\end{figure}

We see similar behavior in the bilayer junctions with $N_b=2$.  The correlation 
strength needed to make an insulating barrier is smaller here, because
the barrier is thicker.  In Fig.~\ref{fig: n=2_f}, we plot the anomalous
average as a function of the plane number.  The planes numbered 31 and 32
are where the barrier lies.  The behavior is like that seen in the
thin barrier case---the correlated metal $(U_{FK}<2.5)$ and pseudogap regions
$(2.5<U_{FK}<5)$ are similar.  The correlated insulator regime, where the
oscillations in the anomalous average increase, but its value within
the barrier is rather insensitive to $U_{FK}$ $(5<U_{FK}<7)$ and the
strong insulating regime $(U_{FK}>7)$ shows an enhancement of the anomalous
average within the barrier at an even larger value of $U_{FK}$ (starting at 
$U_{FK}\approx 14$).

\begin{figure}[htbp]
\epsfxsize=3.0in
\centerline{\epsffile{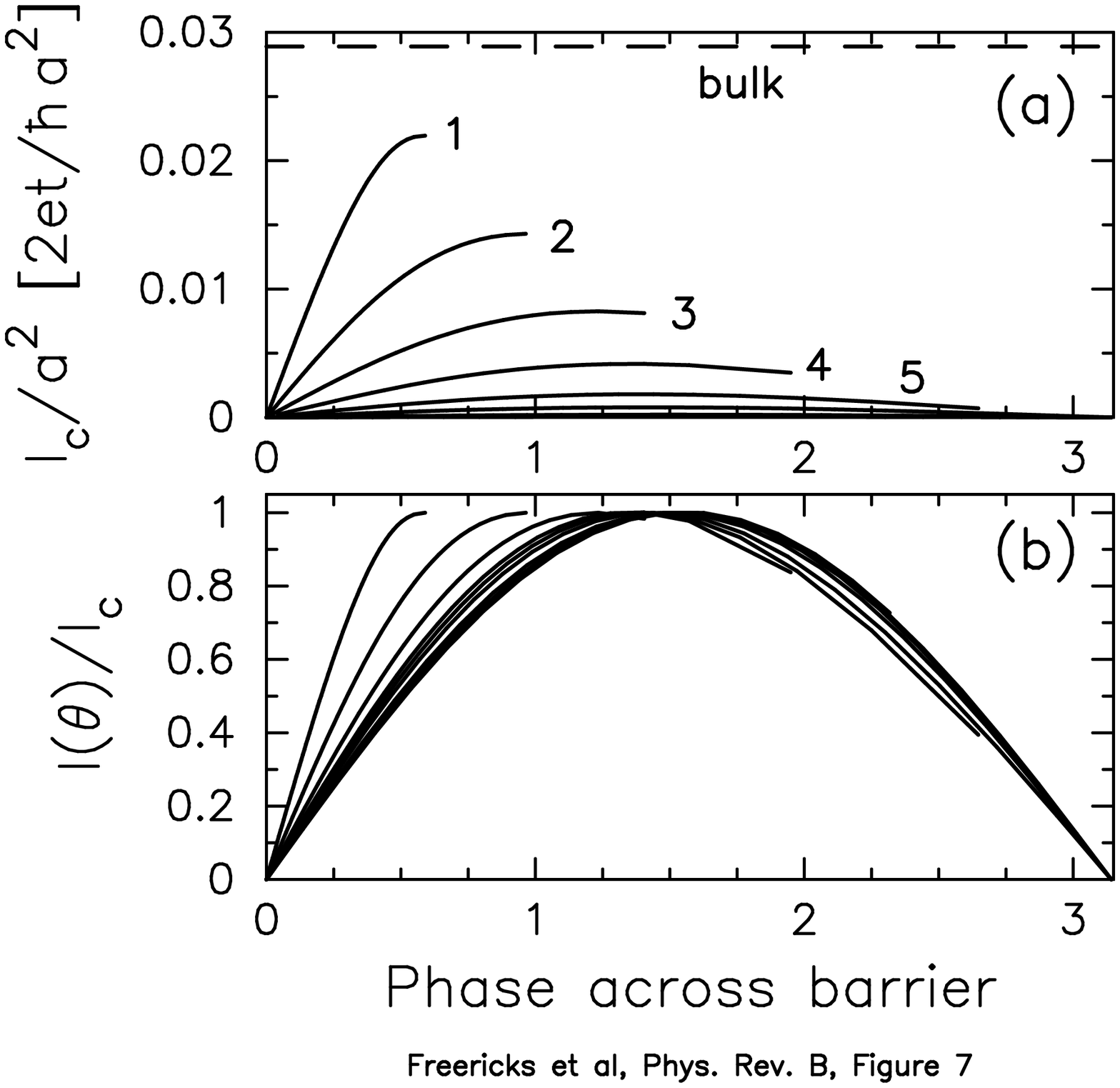}}
\caption{
\label{fig: n=2_current}
Current phase relations for $N_b=2$. (a) Current-phase relation.
Note how the current is typically reduced by about 25\% for an equivalent value
of $U_{FK}$ from the thin barrier result and how the phase difference is 
approximately doubled across the barrier. Values of $U_{FK}$ include
1,2,3,4,5,6,8,10,12, and 16 (labels included for some of the curves).
(b) Normalized current-phase
relation.  Note how the expected sinusoidal dependence enters for 
insulating barriers.}
\end{figure}

In Fig.~\ref{fig: n=2_current}, we show (a) the current versus
phase and (b) the renormalized current-phase relation. The current is
reduced by a factor of about 25\% for an equivalent value of the correlation
strength, and the phase across the junction increases by about a factor
of two versus the single-plane barrier.  As the barrier becomes more 
insulating, we recover the expected 
result that $I(\theta)=I_c\sin\theta$ because now all of the phase
difference takes place over the barrier (in general, the majority of the
phase difference occurs over the central plane of the barrier).  More metallic 
barriers have
the maximum in the $I(\theta)$ curve pushed to values of $\theta$ less than
$\pi/2$ as expected due to the self-consistency and the proximity to the 
bulk critical current.

Self-consistency is an important feature of the calculations.  One might
ask whether the self-consistency modifies the superconductivity much when
current is being carried by the junction versus the zero-current case.
We find that the self-consistency is modified when the barrier is thin
and it is carrying supercurrent.  In particular, we find that the
anomalous average changes as the phase gradient for the bulk system
increases.  For a weakly correlated metal, the largest change is within
the barrier, where the anomalous average decreases by about 30\% as the
critical current is approached $(U_{FK}=2)$.  A smaller decrease is seen in the
superconducting regions close to the barrier.  In the strongly-correlated
insulator regime, the behavior is different---the anomalous average within
the barrier increases by about 10\% as one passes through the full
$I(\theta)$ curve $(U_{FK}=12)$.  The change in the superconducting region is 
much smaller (less than 1\%).  This modification of the self-consistency 
when the junction carries current becomes less important as the barrier
thickness increases---in the single-plane barrier case, we found
the anomalous average within the barrier increases by more than a factor of 2
(for $U_{FK}=12$).

The critical current, normal state resistance and characteristic voltage
appear in Fig.~\ref{fig: n=1_2_icrn}.  The critical current decreases 
much more rapidly for the bilayer than for the single-layer junction.  
The normal state resistance increases more rapidly as well, since the 
bilayer has a resistance that is much more than two times the $N_b=1$
resistance in the strongly insulating limit.
The characteristic voltage is quite interesting, because it has nonmonotonic
behavior.  There is a weak maximum for the moderately correlated
metal (near $U_{FK}=3$), but in the insulating region the voltage 
increases linearly with the correlation strength, attaining values more
than 40\% higher than the Ambegaokar-Baratoff limit.  This is quite
different from what we expect---a constant characteristic voltage---and the
characteristic voltage shows no sign of saturating even at a correlation
energy of $U_{FK}=20$!

\subsection{Moderately thick barrier $(N_b=5)$}

\begin{figure}
\epsfxsize=3.0in
\centerline{\epsffile{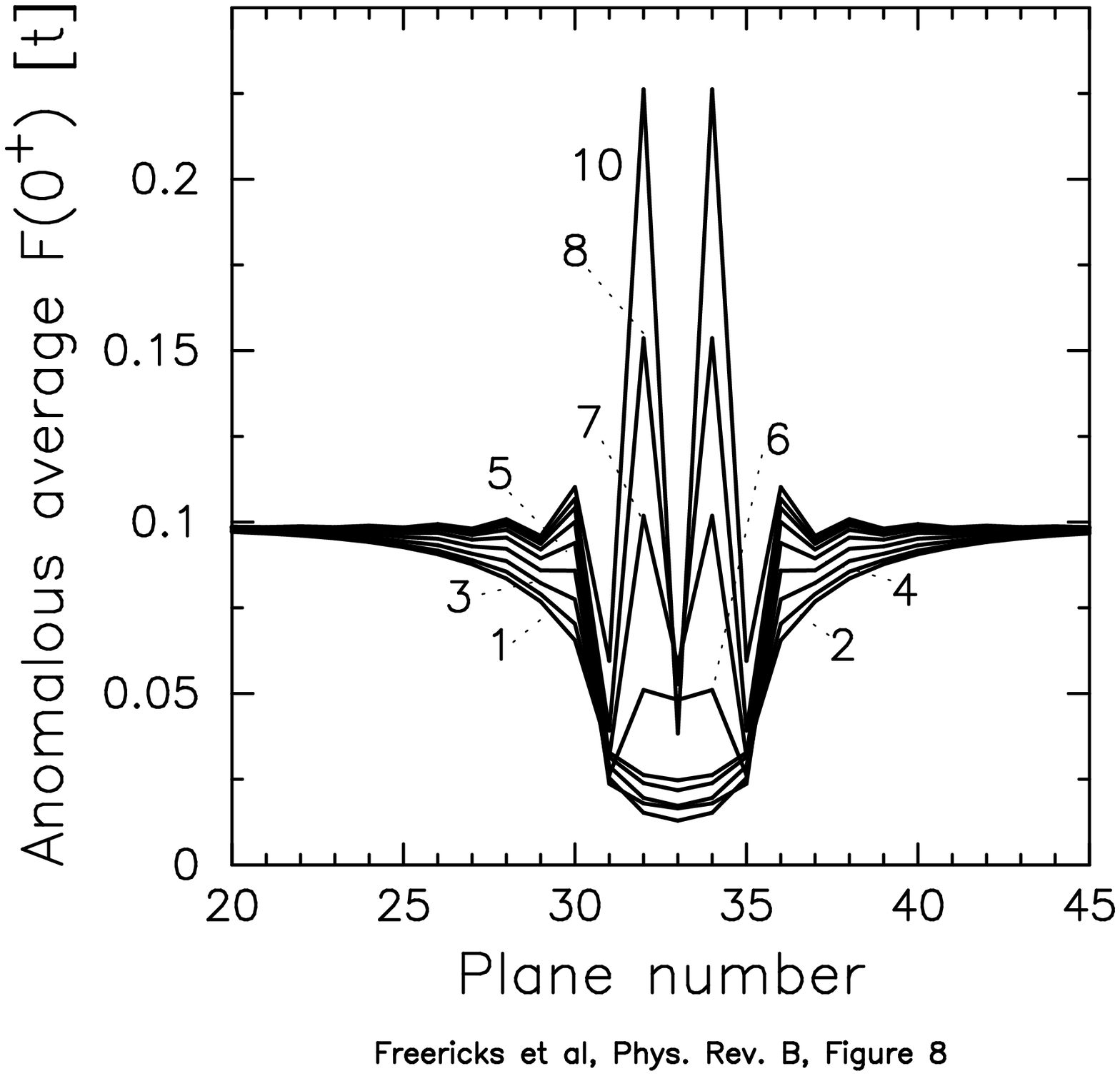}}
\caption{
\label{fig: n=5_f}
Anomalous average plotted versus plane number for the $N_b=5$ junction.
The values of $U_{FK}$ chosen are 1,2,3,4,5,6,7,8, and 10.
Note how the large oscillations are now separated from each other and 
are clearly tied to the superconductor-barrier interface.}
\end{figure}

The barrier region (with $N_b=5$) is chosen to be slightly thicker than the 
bulk coherence length $\xi_0\approx 4a$.  
Here we examine properties of a junction outside of the
regime of most analytic approximations.  In Fig.~\ref{fig: n=5_f},
we plot the anomalous average versus plane number (the barrier lies at
planes numbered 31 to 35).  These results are similar to those
seen before. In the weak correlation regime $(U_{FK}<2.5)$, the anomalous 
average is a smooth function that decreases as the correlations increase.  
Oscillations begin to develop for $2.5<U_{FK}<4.5$, but the anomalous average
continues to decrease within the barrier.  As the correlations increase
further, $U_{FK}>5$, the anomalous average first increases at the center of
the barrier, then a two-peak structure emerges, which has a large amplitude
oscillation and a minimum at the central plane of the barrier.  We can
see clearly here that the oscillatory behavior seen in the previous
cases is arising from effects occurring at the superconductor-barrier
interface as the barrier is tuned through the quantum-critical point
(this is further confirmed with the $N_b=20$ data below).

\begin{figure}
\epsfxsize=3.0in
\centerline{\epsffile{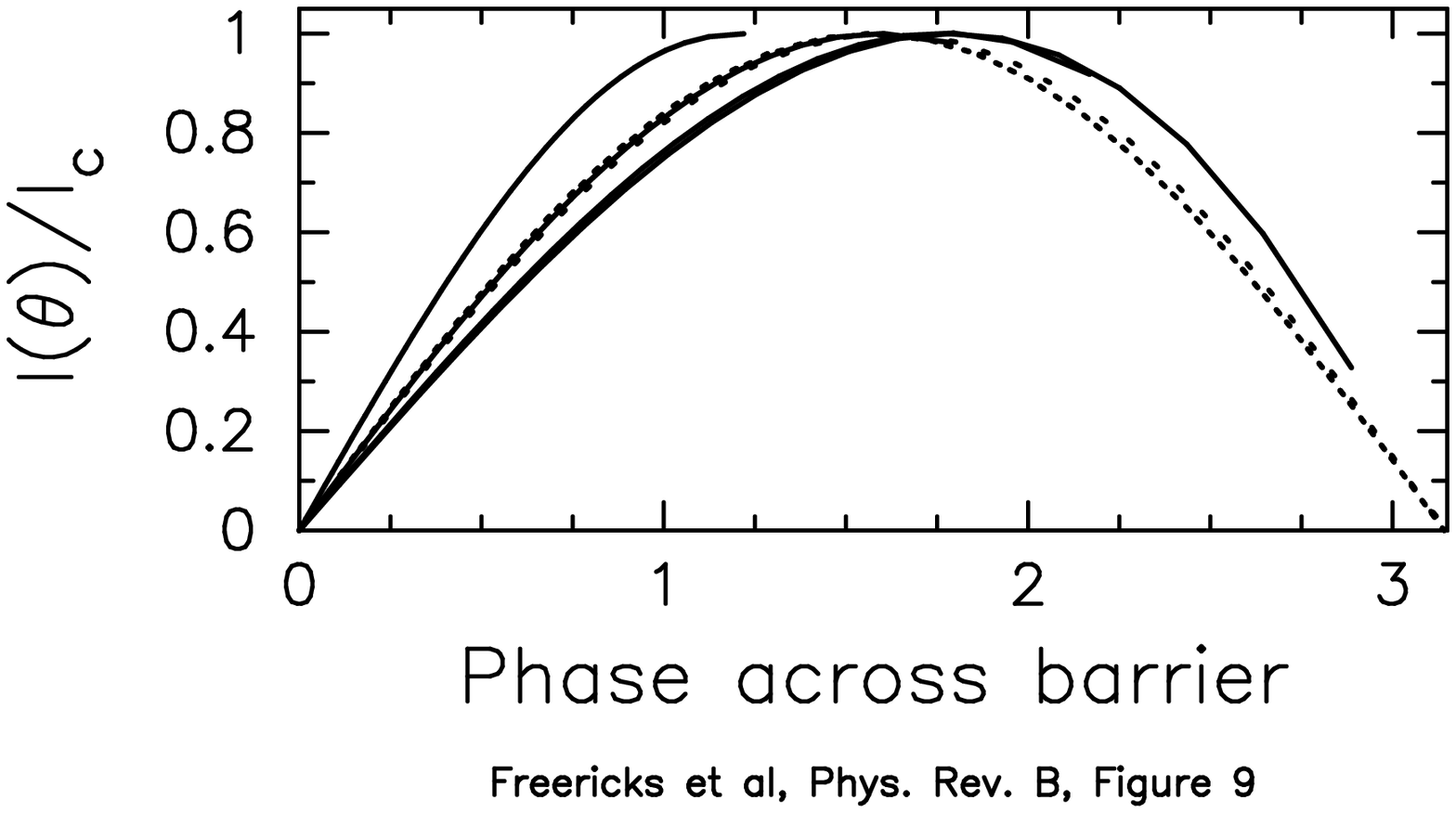}}
\caption{
\label{fig: n=5_current}
Normalized current-phase relation.  Note how the maximum lies at an angle
less than $\pi/2$ for $U_{FK}=1,2$, increases to a value larger than $\pi/2$ for
moderate correlations $U_{FK}=3,4$ and then settles down to $\pi/2$ for larger
correlations $U_{FK}>5$ (curves with a dotted line).}
\end{figure}

The current-phase relation is similar to those seen previously, and will
not be shown here.  The normalized current-phase relation, is plotted
in Fig.~\ref{fig: n=5_current}.  This result is quite interesting.  In the
weakly correlated regime $U_{FK}<2.5$, the maximum of the current-phase relation
occurs at a phase smaller than $\pi/2$ as expected for a thin metallic 
barrier.  As the correlations increase, the maximum first overshoots
$\pi/2$ $(U_{FK}=3,4)$, and then returns to its expected location at $\pi/2$ for
$U_{FK}>5$.  There is a delicate interplay between the strength of the 
correlations and the location of the maximum of the current-phase relation.

\begin{figure}[htbp]
\epsfxsize=3.0in
\centerline{\epsffile{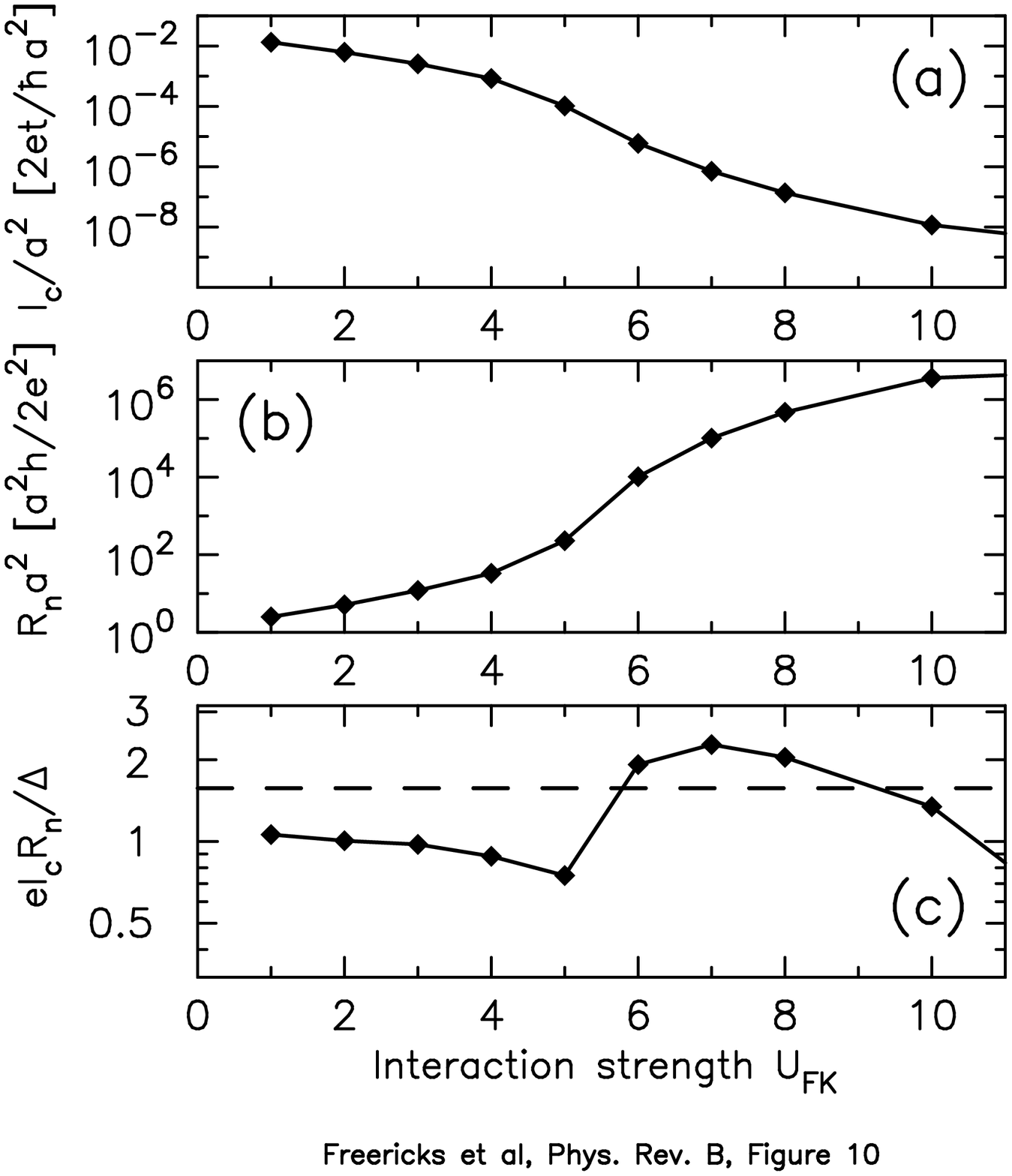}}
\caption{
\label{fig: n=5_icrn}
(a) Critical current, (b) normal state resistance, and (c) characteristic 
voltage
of the $N_b=5$ (diamond) junction. 
The metal-insulator 
transition can be seen in the critical current (a) and  (more easily)
in the resistance (b),
as the regions where the slope of the curves changes most dramatically.  In
the strongly correlated insulating regime, we find the exponential decay
of the current (and increase of the resistance) has a different slope than in
the correlated metal regime. The characteristic voltage (c) has complex
behavior: it first decreases in the metallic regime, then has a sharp
increase at the metal-insulator transition, followed by a decrease as the
correlations increase further (the Ambegaokar-Baratoff prediction is the
dashed line).  Note how the characteristic voltage is maximized just on the
insulating side of the metal-insulator transition, and how the maximal value is
about 40\% larger than the Ambegaokar-Baratoff prediction.  
}
\end{figure}

The critical current, normal-state resistance, and characteristic voltage
are plotted in Fig.~\ref{fig: n=5_icrn} for $N_b=5$ (diamond).  
It is difficult to locate the metal-insulator
transition from the $N_b=5$ critical current data 
(except by focusing on the inflection point), but the transition is clear in
the resistance, which has a sharp increase in the range from $U_{FK}=5$ to 6.
The characteristic voltage has striking behavior.  Starting at a value about 
20\% less than the Ambegaokar-Baratoff limit in the metallic regime, the voltage
initially decreases with correlation strength, then has a sharp increase
(by over 100\%) at the metal-insulator transition, reaching a maximum almost
40\% higher than the Ambegaokar-Baratoff result, until it finally starts to 
decrease as correlations increase further, continuing to decrease at the
largest value of correlations where we performed calculations.  Hence, 
junctions in this regime do
see an enhancement of the characteristic voltage on the insulating side
of the metal-insulator transition.  This behavior is quite complex!

As the barrier size is increased to be on the order of the superconducting
coherence length we see that there is an interplay of a number of different
things: as the barrier becomes more insulating,
oscillations develop in the anomalous average that are pinned to lie near the
superconductor-barrier interface; the supercurrent depends critically on 
self-consistency for metallic barriers, but as correlations increase,
there is an overshoot and the maximum of the current-phase relation occurs
above $\pi/2$ until it settles down to an $I_c\sin\theta$ behavior for
the more insulating barriers; there is little indication of the metal-insulator
transition in the critical current except for the 
appearance of an inflection point (which occurs at the same place that the
current-phase relation has its maximum move to $\pi/2$), but the resistance
shows a clearer picture of the transition. The characteristic voltage is the
most interesting.  Initially it decreases with correlation strength, than
has a sharp increase at the metal-insulator transition, followed by a
maximum and a decrease as correlations are increased further.

\subsection{Thick barrier $(N_b=20)$}

\begin{figure}
\epsfxsize=3.0in
\centerline{\epsffile{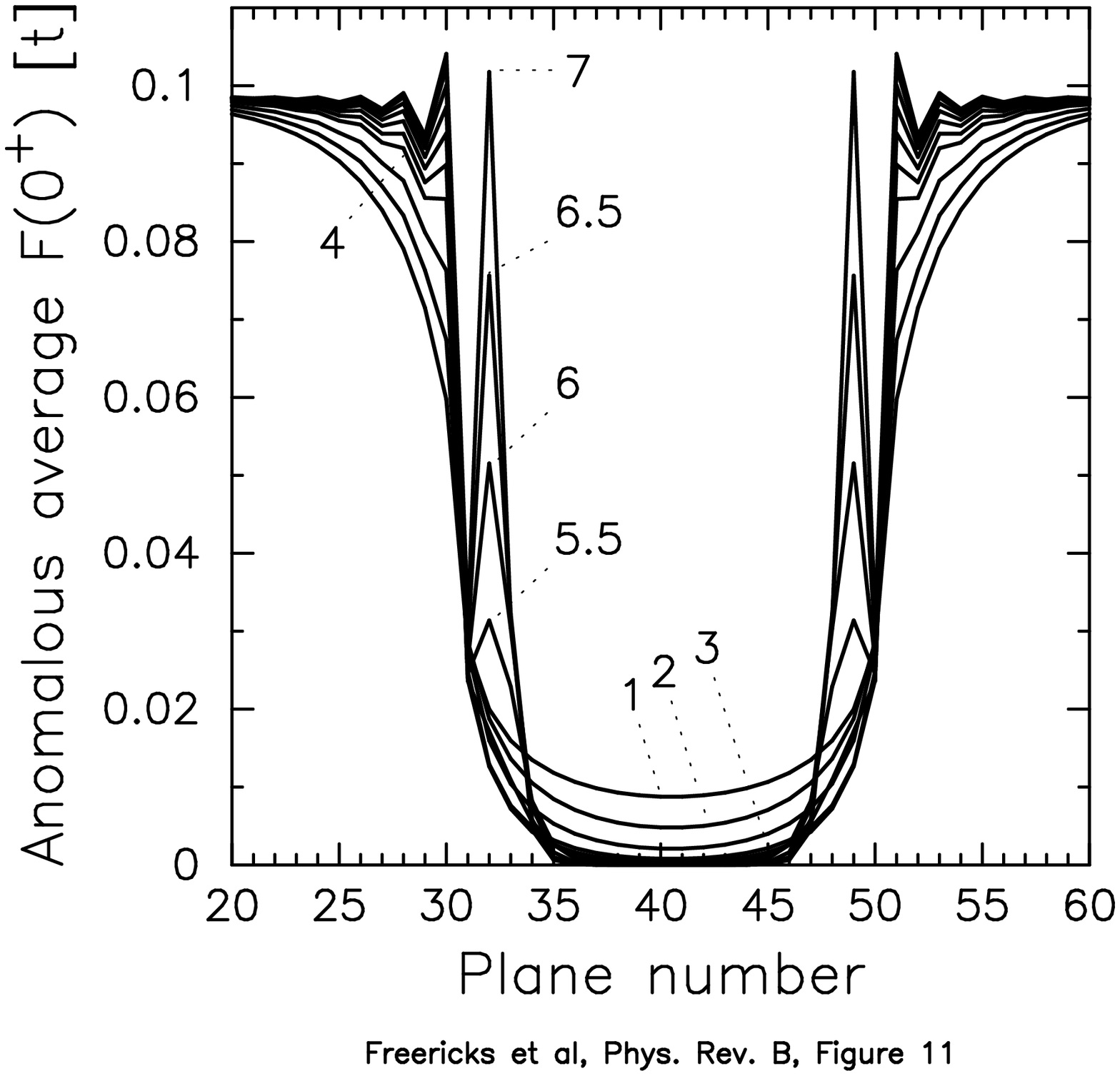}}
\caption{
\label{fig: n=20_f.eps}
Anomalous average for a thick junction $N_b=20$. Values of $U_{FK}$ included are
1,2,3,4,4.5,5,5.5,6,6.5, and 7.  Note how the
metallic regime behaves as expected, but oscillations develop in the
insulating regime that become huge just inside the barrier.}
\end{figure}

The thick barrier junction $N_b=20$ behaves in many respects like the 
bulk barrier material.  The transition from a metal to an insulator 
occurs at approximately $U_{FK}=5$ as in the bulk, and the junction rapidly 
develops
oscillations in the anomalous average at $U_{FK}\approx 4$.  This is shown in
Fig.~\ref{fig: n=20_f.eps}.  Note how the correlated metal regime behaves
entirely as expected---the anomalous average decreases as one approaches and
then enters the barrier, but it never gets too small in the metallic regime
because of the proximity effect.  As the correlations increase, oscillations 
first develop in the superconductor and then move into the interfacial region,
penetrating about one coherence length into the barrier before
they are rapidly suppressed within the barrier.  We see the same phenomenon
as in the thinner junctions: in the insulating regime, the anomalous average
can increase to above it's bulk value within the barrier, but close to
the superconductor-insulator interface.  As the correlations increase, 
the peak of the anomalous average grows.  

The current-phase relation is essentially sinusoidal in these junctions.
For the weakly correlated metals, the peak of the $I(\theta)$ curve
occurs just above $\pi/2$, but as it becomes more insulating, the peak
moves downward toward $\pi/2$ and the current-phase relation becomes 
$I(\theta)=I_c
\sin\theta$, as expected.  There is a computational difficulty that enters
when we are in the strong-insulator regime for a thick-barrier
junction.  Here the critical current
gets exponentially small, and the computational algorithm loses 
current conservation through the entire junction (when the calculation is
halted at a self-consistency error of one part in $10^7$ for the anomalous
average at $\tau=0$).  Instead, we see an
exponential decrease of the current from the value fixed at the
bulk superconductor to the value within the barrier.  The current is
constant within the barrier itself, and the current-phase relation is
a nice sine curve, so we believe that the critical current found from
our algorithm is accurate, even though, the boundary conditions with the
bulk are trying to force more current through the junction than it can
have; i.e., the current discontinuity occurs far from the barrier
region.  This scenario is similar to that of Josephson's original
analytic scheme\cite{Jose}, since in his case, there is no phase gradient over 
the superconductors, so they carry no current, but there is current in the
barrier, since there is a phase difference across it.

\begin{figure}[htbp]
\epsfxsize=3.0in
\centerline{\epsffile{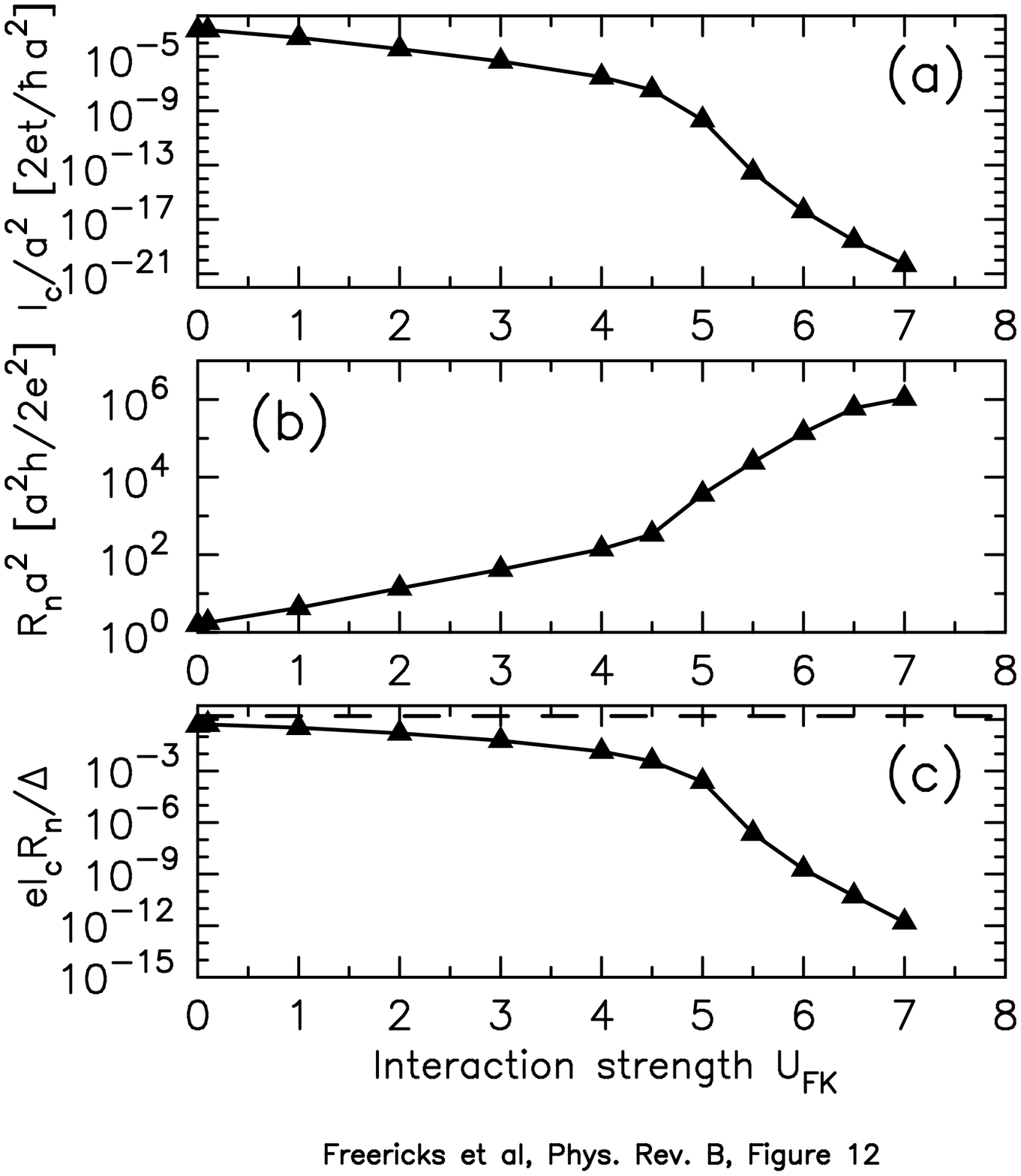}}
\caption{
\label{fig: n=20_icrn}
(a) Critical current, (b) normal state resistance, and (c) characteristic
voltage
of the $N_b=20$ (triangle) junctions.  Note how the
thick junction behaves much like the bulk material.  The metal-insulator
transition can be clearly seen in the critical current and in the resistance,
as the regions where the slope of the curves changes most dramatically.  In
the strongly correlated insulating regime, we find the exponential decay
of the current (and increase of the resistance) has a different slope than in
the correlated metal regime.  Note that the thick barrier has a sharply
suppressed characteristic voltage because the decrease in $I_c$ is much
sharper than the increase in $R_n$. We believe this occurs because the
temperature dependence of the resistance is strong, even at these low
temperatures.  The dashed line is the Ambegaokar-Baratoff prediction.}
\end{figure} 

The critical current, normal-state resistance, and characteristic voltage
appear in Fig.~\ref{fig: n=20_icrn}.  One can see the metal-insulator
transition clearly in the $I_c$ curve.  Within the metal, the critical
current has an exponential dependence on $U_{FK}$, within the insulator it has
a different exponential dependence.  In the transition region, it 
decreases most sharply.  We believe the reduction in the rate of decrease in
the critical current as $U_{FK}$ is increased into the correlated insulator 
regime arises in part
from the fact that the effective junction thickness is thinner than
the true junction thickness due to the oscillations in the anomalous average
that develop within the barrier region.  The resistance shows the expected
behavior as well.  One can clearly see the metal-insulator transition as the
region where the conductivity changes it's functional dependence sharply.
Note, however, that the characteristic voltage is severely affected by the
metal-insulator transition, since the decrease in the critical current far 
outweighs the increase in resistance, and the characteristic voltage decreases
by many orders of magnitude as one enters the strongly-correlated insulator.
One reason why this occurs is because the resistance for the correlated
insulator depends strongly on temperature.  As $T\rightarrow 0$, the resistance
becomes very large, but it can be sharply reduced
as the temperature increases.  Hence, even though the critical current is at
the zero-temperature limit for $T/T_c=0.1$, the resistance still has strong
temperature dependence in this regime, which causes the characteristic
voltage to be sharply suppressed.  Perhaps this behavior plays a role in
some junctions that appear to work well at low temperatures, but then fail
as the temperature is increased, because on the insulating side of the
metal-insulator transition, the resistance has strong temperature
dependence.

\subsection{Summary of tuning the correlation energy}

We have discovered a number of interesting features of Josephson junctions
for short-coherence-length superconductors that have their barrier
tuned through the quantum-critical point of
a metal-insulator transition.  The most striking feature we
find is that in the insulating regime, there are oscillations with a 
wavelength on the order of the Fermi wavelength, that appear at the
superconductor-barrier interface and decay on the order of the coherence
length on either side of the interface.  They can have very large amplitudes
(on the order of the bulk value of the anomalous average) within the
barrier.  We believe that these oscillations are occurring from a ``surface''
effect intrinsic to the superconductor, and depending on how close the
two interfaces are (determined by the thickness of the barrier) they are
either independent of each other or can interfere.  Note that these results
differ from those found in metallic junctions with ``geometrically diluted''
barriers\cite{levy1}.  There, oscillatory behavior was seen even for metallic
barriers $U_{FK}=0$, and the decay length was much longer, leading to a number
of cycles before the oscillations are damped. We also found interesting
results for the current-phase relations.  As expected, thin junctions 
typically have $I_c$ occur at a phase difference smaller than $\pi/2$, but
in all but the single-plane junction, as the correlations increase, the
maximal $I_c$ occurs at $\pi/2$ and the curve becomes sinusoidal.  For thick
barriers, we find the maximum occurs larger than, but close to $\pi/2$ for
metallic barriers and then migrates towards $\pi/2$ as the barrier becomes
more insulating.  Finally, we found new behavior in the characteristic
voltage of a junction. The characteristic voltage is limited in the metallic
regime by the bulk critical current of the superconductor multiplied by
the junction resistance for a clean barrier (the so-called ``planar contact''
limit).  This value is approximately
$1.31\Delta /e$, which is a about 8\% smaller than the 
Ambegaokar-Baratoff result for an insulating barrier.  As the correlations
increase, $I_c$ decreases to be much below the bulk critical current of
the junction, and $R_n$ increases.   The characteristic voltage has
a rich behavior.  For the thin junction $(N_b=1$)
it is maximized in the correlated
metallic regime, and becomes constant for the insulator.  As the thickness 
increases to $N_b=2$,
we continue to see a small maximum in the metallic regime, but the
interesting behavior is that for a wide range of correlation strengths, the
Ambegaokar-Baratoff result does not hold, and the characteristic voltage
increases with correlation strength.  For barrier thicknesses on the order of
the correlation length ($N_b=5$), the behavior is even more complex. The voltage
initially decreases with correlation strength, then has a sharp rise at
the metal-insulator transition, followed by a maximum for the correlated
insulator that ultimately decreases as the correlations increase further.
The Ambegaokar-Baratoff regime doesn't hold here either.  Finally
in the thick junction regime $(N_b=20$), the resistance has a strong dependence 
on temperature in the insulating regime, and even at what is a low temperature
for the superconducting properties, the characteristic voltage can decrease
significantly as the correlations increase. The conclusion that can be
drawn from this is that one requires a careful tuning of the thickness
of the barrier, the proximity to the metal-insulator transition, and the
operating temperature to optimize the properties of a junction.
This idea is further supported in the next section.

\section{Tuning the junction thickness through the metal-insulator transition}
\label{sec:tuning_thick}

Here we present results on tuning the junction from the thin to
thick barrier at three values of $U_{FK}$: (i) $U_{FK}=2$ a weakly
correlated metal; (ii) $U_{FK}=4$ a strongly correlated
(pseudogap) metal; and (iii) $U_{FK}=6$ a correlated insulator.
In the correlated metal case, the junction can be viewed as an SNS
weak link,\cite{likharev_review} while the correlated 
insulator barrier eventually leads to an SIS junction. 
In the case with $U_{FK}=2$, the normal region is a non-Fermi liquid metal
that can be described as dirty metal (resistivity $\rho_n
\simeq 240\ \mu \Omega$cm with the assumption that the lattice constant
is 3 \AA), for $U_{FK}=4$, we get a ``bad metal''
($\rho_n \simeq 2700 \ \mu \Omega$cm; such huge resistivities do not
necessarily require electronic correlations,\cite{gunarson} but
are also seen in model calculations involving disordered
Fermi liquids\cite{allen}). The early experimental\cite{snsexp}
and theoretical\cite{degennes} work on SNS junctions revealed
that the supercurrent in these structures arises from the proximity
effect: superconducting correlations are generated  in the normal
region with the density of pairs decreasing exponentially from the SN
interface on a scale set by normal metal coherence length
$\xi_n=(\hbar {\mathcal D}/2\pi k_BT)^{1/2}$ (here, ${\mathcal D}$ is a classical 
diffusion constant). The equilibrium current then flows at zero voltage
because of the overlap of pair-field wave functions from the two
superconductors. Recent mesoscopic advances have supplemented this 
``crude'' picture with the analysis of energy-resolved 
quantities~\cite{schon,yip2} which become important for phenomena 
on small length scales at low temperatures and voltages.~\cite{french} 
The initial theoretical studies\cite{degennes,fink}
relied on Ginzburg-Landau theory (which formally requires $T$ to
be close to $T_c$) in the dirty limit ($\xi_0 \gg \ell$, with $\ell$
the mean free
path) and for long junctions ($N_b \gg \xi_n$)\cite{degennes}. In the ensuing
approaches, based on quasiclassical Green's function formalism,
junctions with more general parameters were 
described,\cite{likharev_review,zaikin} 
where the proximity effect on the superconducting side (i.e., a depression 
of the order parameter near the SN interface) was taken into 
account\cite{kupriyanov} (such effects are treated from the onset
in self-consistent studies like ours). Thus, the conventional 
proximity effect theories show that the critical current is 
determined primarily by the behavior of the superconducting 
order parameter when crossing an SN boundary, while its thickness dependence 
and temperature dependence are affected by the way quantum 
coherence is  lost in the normal metal. However, it is only recently 
that mesoscopic studies\cite{schon2} have emphasized the importance 
of the Thouless energy $E_{\rm Th}=\hbar {\mathcal D}/N_b^2=2\pi k_B 
T\xi_n^2/N_b^2$ for the proximity effect\cite{schon}. 
Although $E_{\rm Th}$ is determined by the classical diffusion 
time $N_b^2/{\mathcal D}$ for a particle to cross the sample, it 
plays a prominent role in various quantum phenomena 
encountered in disordered (normal) metal electron physics.\cite{janssen} 
In the long junction limit $\Delta \gg E_{\rm Th}$, 
the critical current is set by $E_{\rm Th}$---according 
to the recent quasiclassical (non-self consistent) 
calculations\cite{schon2} $eI_cR_n(T=0)=10.82E_{\rm Th}$.  
In the short junction limit $E_{\rm Th} \gg \Delta$,
for $T \rightarrow 0$, the product 
$I_cR_n$ is expected to be given by the diffusive 
limit $0.66\pi\Delta/e$ of the Kulik-Omelyanchuk 
formula\cite{kulik}, or  $1.92 \Delta/e$ in the case of 
dirty interface with Schep-Bauer distribution of 
transparencies\cite{bardas,yehuda} 
(as discussed in Sec.~\ref{sec:tuning_ufk}). The high 
versus low temperature limit is set\cite{schon2} by 
the ratio of $k_BT$ and $E_{\rm Th}$, or, equivalently, 
$N_b$ and $\xi_n$, since $N_b=\xi_n$ is defined to be the length 
scale at which $k_BT=E_{\rm Th}$.

While the energy gap $\Delta$ is determined by the (attractive) 
electron-electron interaction in the superconducting leads,  $E_{\rm Th}$ 
is a single-electron concept, and as such is not directly applicable to 
our correlated metal (which has no well-defined Landau quasiparticles). 
Nevertheless, it is a common practice 
in experimental studies\cite{htc} to extract estimates for such 
``quasiparticle'' parameters\cite{htc} using measured values of $\rho_n$ 
(even for unusually high resistance barrier materials, like the 
underdoped cuprates\cite{htc}), and check if the conventional treatment 
can describe the junction.  From the formal point of view this corresponds 
to comparing the correlated junction to an SNS weak link with a well understood 
normal metal in the barrier region (described in terms of non-interacting 
quasiparticles), which has the same resistivity as the real 
barrier material [in fact, most of theoretical treatments operate with simple,
usually phenomenological, description of the barrier region in terms of 
its transmission properties $D$ which reproduce the measured 
average resistance $1/R_n=\int_0^1dD P(D) D$, as discussed in 
Sec.~\ref{sec:tuning_ufk}]. 
Therefore, for the sake of comparison of our result with standard calculations,
we extract a diffusion constant ${\mathcal D}$ from the Einstein relation
$1/\rho_n=2e^2N(E_F){\mathcal D}$ [with $N(E_F)$ the (single-spin) interacting 
density of states at the Fermi energy]. This  is independent of  
band-structure effects (classically ${\mathcal D}=v_F\ell/3$,
but ${\mathcal D}$ can also be defined quantum-mechanically 
from the Kubo formula in an exact state representation,\cite{allen} 
which then allows one to use a diffusivity even when the semiclassical 
picture of the mean free path $\ell$ breaks down). For the dirty-metal 
case $U_{FK}=2$ we find ${\mathcal D}_{U_{FK}=2} \approx 2 ta^2/\hbar$ 
and for the bad-metal case we find ${\mathcal D}_{U_{FK}=4} \approx 0.32 
ta^2/\hbar$. The corresponding normal metal coherence lengths are 
$\xi_n \approx 5.6a$ and $\xi_n \approx 2.3a$ (at $T=0.01$) in the former 
and latter case, respectively. 

\begin{figure}
\epsfxsize=3.0in
\centerline{\epsffile{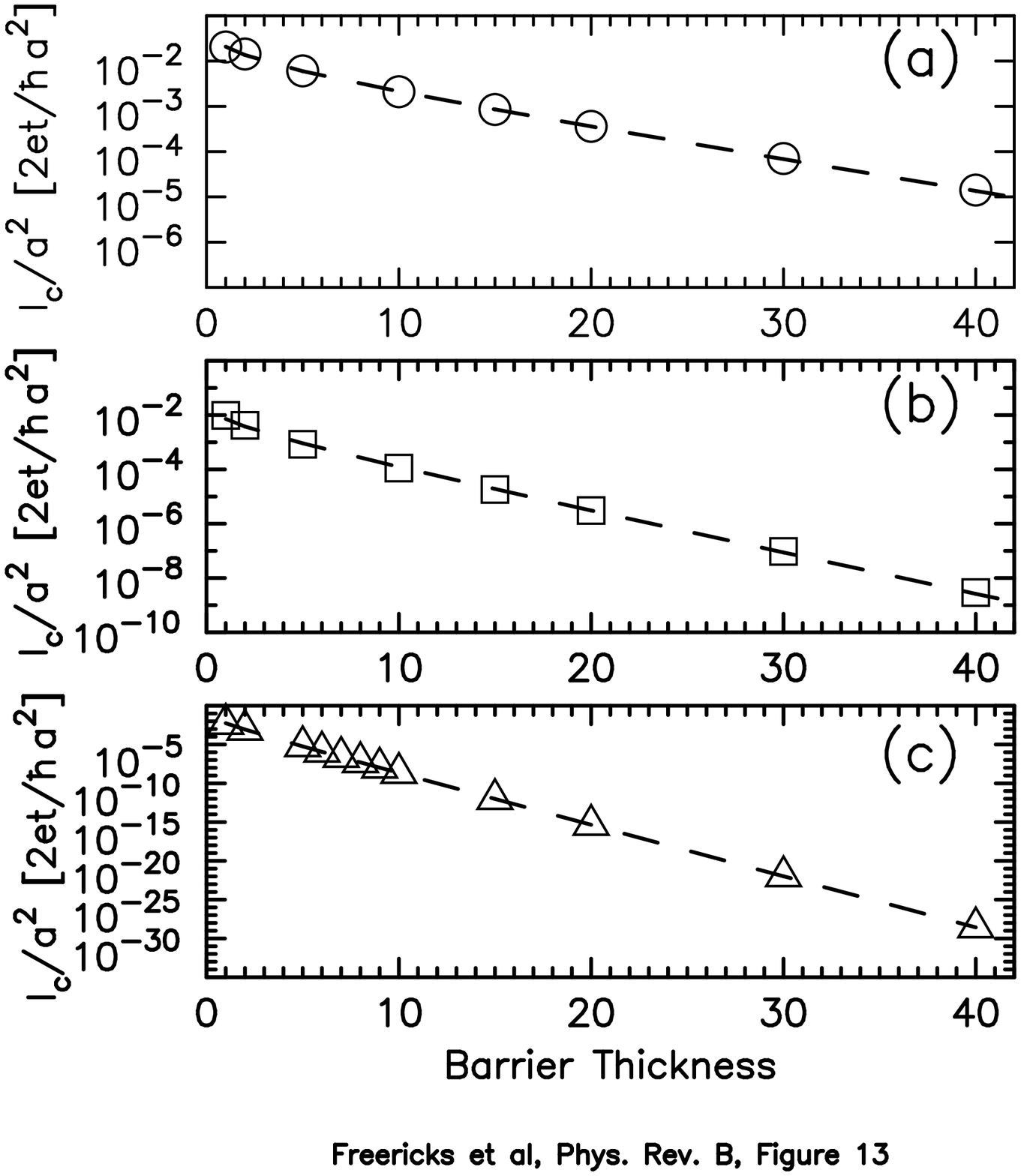}}
\caption{
\label{fig: thickness_ic}
Critical current versus the barrier thickness for (a) a weakly correlated
metal $(U_{FK}=2,$ circle), (b) a strongly correlated metal $(U_{FK}=4$, 
square), and  (c) a
correlated insulator near the quantum-critical point $(U_{FK}=6$, triangle).  
Note how the shapes of these curves vary as the correlations increase.
The dashed curve is a fit of the data  for $N_b\ge 5$ to the form given in 
Eq.~(\ref{eq: ic_thickness}).
}
\end{figure}

The critical current is plotted in a semilogarithmic plot in
Fig.~\ref{fig: thickness_ic}.  The symbols are the calculated values and
the dashed lines are a fit to the following form
\begin{equation}
I_c=A N_b^x \exp [- N_b/\xi_b],
\label{eq: ic_thickness}
\end{equation}
with $A$ a constant, $\xi_b$ the coherence length in the barrier (the symbol 
$\xi_b$
is used here to differentiate it from the phenomonological $\xi_n$ determined
from a diffusive metal analogy above),
and $x$ an exponent (we only fit the data with 
$N_b\ge 10$, since the thin-plane results are sufficiently different from the 
thick-plane results, and the fits are therefore much more accurate;
nevertheless, the final functional form for the data works well for
all barrier thicknesses).  We find that the fits
vary from the analytic forms for the thick barrier
limit ($x=1$).  For example we find that the coherence length decreases from
$\xi_b=6.66$ for $U_{FK}=2$, to
$\xi_b=2.96$ for $U_{FK}=4$, to $\xi_b=0.665$ for $U_{FK}=6$.  Similarly,
the exponent varies from $x=-0.40$
for $U_{FK}=2$, to $x=-0.45$ for $U_{FK}=4$, to $x=-0.53$ for
$U_{FK}=6$.  The value for the exponent never becomes
close to the asymptotic result of $x=1$ for a thick junction. But the
coherence length behaves as expected---as the scattering increases, the
coherence length decreases, becoming very small as the barrier goes
through the metal-insulator transition and becomes a correlated insulator
(in fact, our values for $\xi_b$ are only about 20\% larger than
the estimates for $\xi_n$ given above).
Note that this fitting procedure is not well-defined, 
since we do not have data over many decades of barrier thicknesses and 
because we can trade-off some effects of the fitting by simultaneously
changing the
exponent and the coherence length. But in all cases shown, we fit all
of the data from $N_b=5$ to $N_b=80$ to an accuracy of better than 10\% for
the critical current (the accuracy decreases to about 25\% for $N_b=1$).
This fitting scheme with nontrivial exponents is definitely more accurate than
the best fits one can achieve with $x=1$ or $x=0$. We find the case with 
$U_{FK}=4$ to be the toughest case to fit to this form (the spread in error
is about 10\% here, with 25\% error for $N_b=1$), while the $U_{FK}=2$
is the easiest, with a fit for all values $N_b=1$ to 80 being accurate to
5\%. 

\begin{figure}
\epsfxsize=3.0in
\centerline{\epsffile{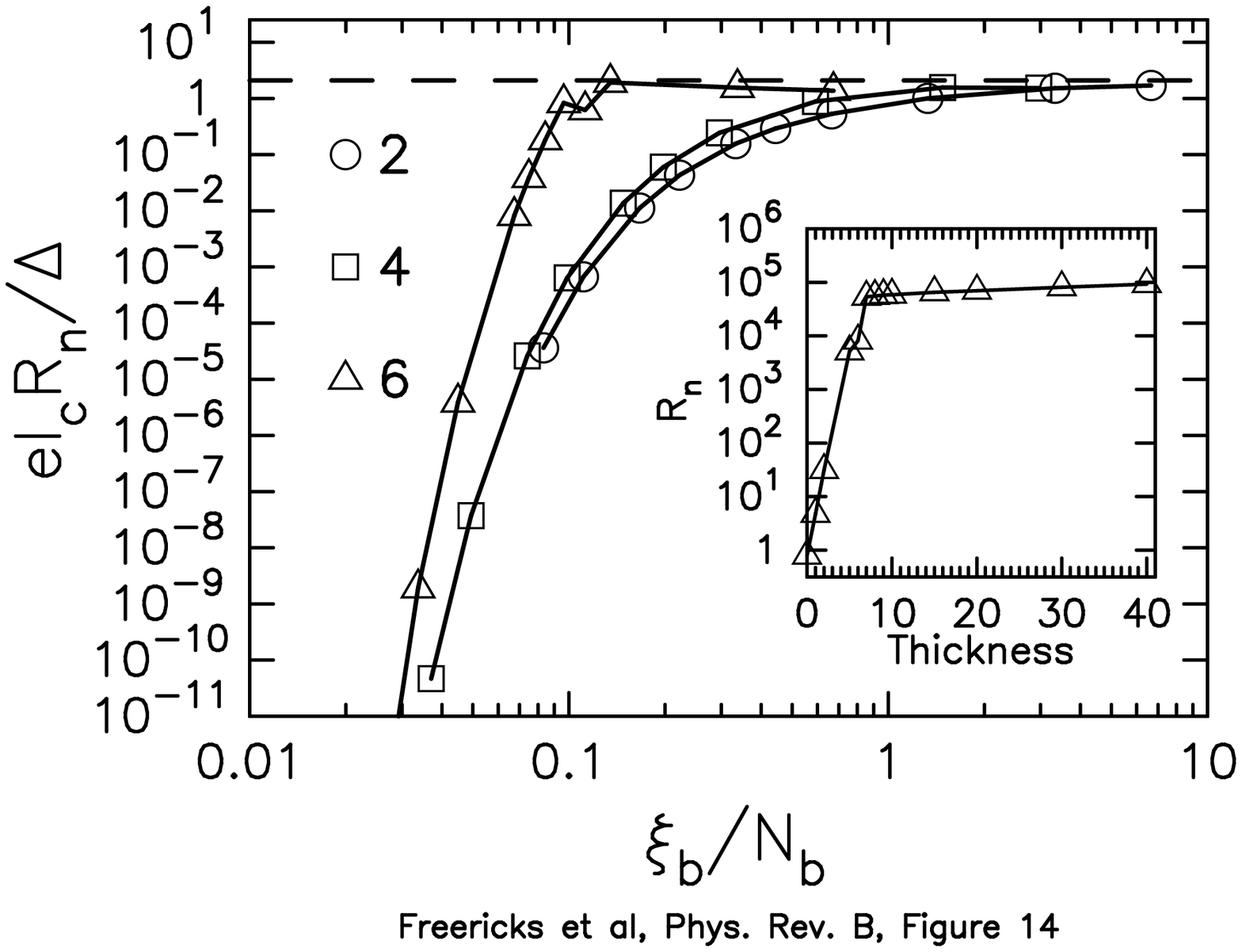}}
\caption{
\label{fig: thickness_icrn}
Characteristic voltage plotted versus the inverse of the effective 
thickness of the barrier on a log-log plot.  Using the correlation length 
extracted from 
Fig.~\ref{fig: thickness_ic}, allows us to plot the characteristic 
voltage against a measure of the Thouless energy $E_{\rm Th}
=2\pi k_BT\xi_b^2/N_b^2$.  
Such a plot should show scaling behavior, according to the quasiclassical 
theory; we find this to be approximately true for the
metallic junctions ($U_{FK}=2$, circles; and $U_{FK}=4$, squares),  but
the correlated insulating barrier has a much sharper dependence on
the barrier thickness (including an ``oscillation'')
and the scaling of the quasiclassical theory
breaks down.  Note the sharp onset of
insulating behavior at a thickness $N_b\approx 7$ for $U_{FK}=6$.
Inset is the resistance versus barrier thickness for $U_{FK}=6$.  Note
the sharp location of the metal-insulator transition near $N_b=7$.
}
\end{figure}

In Fig.~\ref{fig: thickness_icrn}, we plot the characteristic voltage
(in units of $\Delta/e$) versus the ratio of the barrier coherence length
(determined from Fig.~\ref{fig: thickness_ic} and
equivalent to the Thouless length) to the barrier thickness. This is
our analogue of the recent results of a quasiclassical theory\cite{schon2},
which show deviations of the Kulik-Omelyanchuk relation for long
diffusive junctions.  The results for metallic junctions $U_{FK}=2,4$,
have the same shape as seen in the quasiclassical theory, and they nearly scale
with each other (the scaling could be improved by slightly changing
the barrier coherence length).  The correlated-insulator results,
$U_{FK}=6$, however, show a different functional shape, with
the transition from the nearly constant characteristic voltage to the
region where it decreases sharply, occurring much more rapidly than in the
metallic case (and having a small ``oscillation'' at the ``transition'').  
One can be more quantitative in the comparison with the quasiclassical
predictions: in the long-junction limit, the characteristic voltage
is predicted\cite{schon2} to behave like
\begin{equation}
I_cR_n=A^{\prime}N_b^{x^{\prime}}\exp[-N_b/\xi_b],
\label{eq: icrn_quasi}
\end{equation}
with the coherence length determined from the functional dependence of the
critical current on $N_b$ in Eq.~(\ref{eq: ic_thickness}). The quasiclassical
prediction gives $A^{\prime} \approx 5.49 {\mathcal D}/\xi_b^2$ and $x^{\prime}=1$.  
While we can fit reasonably well to this functional form in the regime
where $\xi_b/N_b<1$, we typically find the constant $A^{\prime}$ is 
about three to five times larger and the exponent $x^{\prime}$ is about a 
factor of two
smaller than the quasiclassical predictions for the correlated metal cases
$(U_{FK}=2$ or 4).  The
parameters deviate significantly for the correlated-insulator phase (where the
fitting breaks down severely for $\xi_b/N_b>0.1$).
{\it This shows that the correlated-insulator regime cannot
be described by the conventional quasiclassical approach.} There appears to be 
a critical length at which point the characteristic voltage changes from an
essentially constant
dependence on the barrier thickness to a rapidly decreasing dependence on the
thickness (which is $N_b\approx 7$ for $U_{FK}=6$). The difference in shapes 
seen in Fig.~\ref{fig: thickness_icrn} arises mainly from the behavior of the 
resistance, which assumes a linear scaling with the thickness $N_b$ in the
metallic regime and in the thick insulating regime (although it has an 
additional constant there, when extrapolated to $N_b=0$), but has a rapid
crossover to the linear regime for the thin insulator (semilogarithmic plot 
shown in the inset to Fig.~\ref{fig: thickness_icrn}).

One may wish to conclude from Fig.~\ref{fig: thickness_icrn} that correlated 
insulating barriers are superior to metallic barriers since the parameter
$\xi_b/N_b$ can be reduced to much smaller values than in the metallic
cases before the characteristic voltage becomes reduced.  But such a view 
is erroneous, because the significantly smaller values of $\xi_b$ for the
insulating barriers means that the barrier thicknesses 
where the characteristic voltage starts to
decrease are indeed smaller for the correlated insulator.  What can be inferred
from the figure, however, is that once one reaches the critical thickness where 
the barrier has a metal-insulator ``transition,'' the characteristic voltage is
very strongly dependent on the thickness of the junction.  Hence variations
in the thickness of the barrier can have a large effect on the performance of 
a junction with a correlated-insulator barrier.  In particular, variations
in the thickness could make junctions appear to have ``pinholes'' because
slightly thinner areas can have greatly enhanced Josephson coupling. This
can possibly explain why it appears to be more difficult to attain
small spreads in junction properties for high-$T_c$-based junctions, even
if the barrier is pinhole free because the proximity to the thickness
triggered metal-insulator transition generates ``intrinsic pinholes'' within 
the correlated insulator.

\begin{figure}
\epsfxsize=3.0in
\centerline{\epsffile{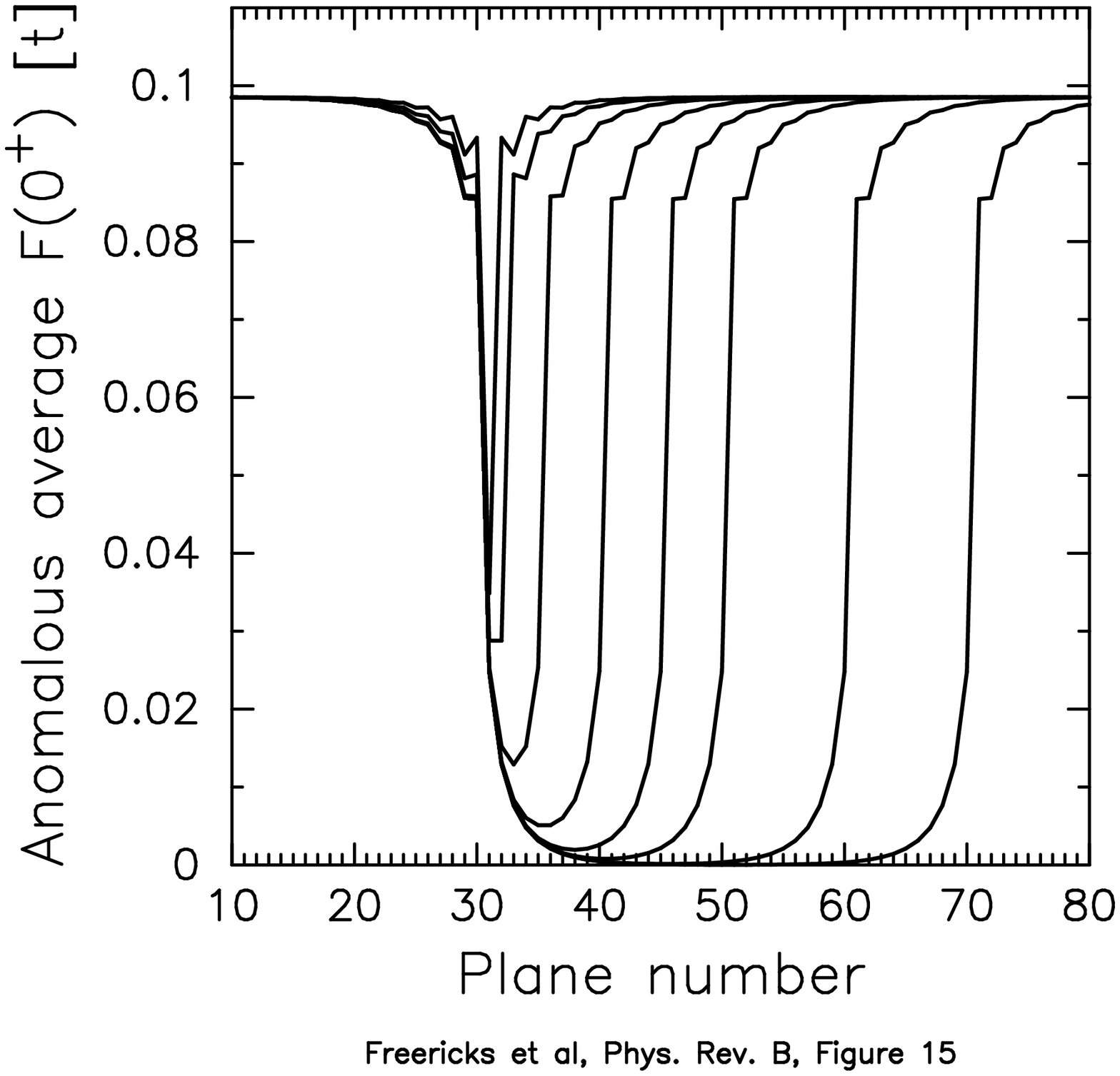}}
\caption{
\label{fig: f_thick}
Anomalous average versus plane number for $(U_{FK}=4)$ and thicknesses
ranging from $N_b$=1,2,5,10,15,20,30, and 40 (the thickness of each 
barrier is obvious from the range of the plots).  Note
how the shapes of these curves are identical for the regions close to the
superconductor-barrier interface  at planes 25--34
(and within the superconducting region to
the right or the left for the right or left interface, respectively).  Since
this shape stops changing after $N_b=2$, we conclude that the oscillations 
are an intrinsic property of the bulk superconductor terminated on 
the barrier.
}
\end{figure}

It is also interesting to examine how the anomalous average behaves as a 
function of the thickness of the barrier as well.  We find the following
result shown for $U_{FK}=4$ in Fig.~\ref{fig: f_thick}: once the 
thickness is larger than the bulk coherence
length (i.e. for all barriers simulated with $N_b\ge 5$), we find that
the shape of the anomalous average is identical for all thicknesses
for the planes that lie within the superconducting region and that
penetrate two to three planes into the barrier.  What this tells us is
that the thickness of the barrier is not influencing the shape of
the anomalous average except within the barrier itself, so the
oscillations are a property of the bulk superconductor 
coming in contact with the barrier.  These results are also true for the
$U_{FK}=2$ and $U_{FK}=6$ cases, but we don't show those results here,
because the agreement is essentially the same as seen in the $U_{FK}=4$
figure below.

Finally, we examine the current-phase relation.  We find similar
behavior to that found when one fixed the thickness and tuned the
correlation strength through the metal-insulator transition.
For the weakly correlated metal $(U_{FK}=2)$, we find that the maximum of
the current-phase relation occurs for phase differences much less than
$\pi/2$ as expected.  As the barrier is made thicker, the maximum first
overshoots $\pi/2$ and then as $N_b> 20$, it settles down at $\pi/2$.
This is also seen in the pseudogap phase $U_{FK}=4$, but 
the spread in values for the maximum of the current-phase relation 
remains clustered closer to $\pi/2$.  By the time we reach the insulator
phase $U_{FK}=6$, the maximum monotonically increases from below $\pi/2$
for thin barriers to $\pi/2$ for thick barriers (with no overshoot).

\section{Conclusions}

In this work we have examined what happens as the barrier of a junction
is tuned from a metal to an insulator for short coherence-length
$s$-wave superconductors.  We studied the transition both as a function of
the correlation strength and of the barrier thickness.  We found a 
number of interesting results.  First, in regimes where the critical 
current density approaches that of the bulk superconductor, self-consistency
is important in determining the current-phase relation, and it is modified 
dramatically from simple sinusoidal behavior.  As the correlations increase,
and the current density decreases, the sinusoidal behavior is restored, but
in some cases, the maximum of the current-phase curve overshoots $\pi/2$ and
then becomes sinusoidal only at an even
larger correlation strength.  Second, we found that
as the barrier becomes more insulating, the anomalous pair-field average
develops oscillations on the order of the Fermi wavelength, which can
be quite substantial in amplitude (up to about twice the bulk anomalous 
average). These oscillations are tied to the superconductor-insulator
interface, and depend little on the thickness of the barrier once the
thickness is larger than about twice the bulk coherence length.  Third,
we found that the critical current has a nontrivial dependence on the 
thickness of the barrier---while it decays exponentially with thickness,
it also has a power-law prefactor that varies with correlation strength,
and deviates sharply from the quasiclassical prediction.
The barrier coherence length decreases, of course, as the correlation
strength increases into the insulating regime.  Fourth, we found that
the characteristic voltage has rich behavior.  It
is maximized for weakly correlated metallic barriers
for thin junctions and the Ambegaokar-Baratoff result is recovered at
strong correlations.  As the barrier thickness increases, the maximum in
the metallic region is reduced, but the Ambegaokar-Baratoff result fails
as the correlations increase, with the voltage increasing linearly 
with $U_{FK}$ over a wide range of correlation strengths.  The intermediate
thickness junctions have the most interesting behavior---the voltage initially
decreases, has a sharp increase at the metal-insulator transition, and then
decreases in the large correlation limit.
Thick insulating barriers have very low characteristic voltages  and strong
temperature dependence, as expected at finite temperatures, since the
junction resistance decreases rapidly as the temperature is increased
in the insulating regime.
We also saw that self-consistency can renormalize the Ambegaokar-Baratoff
limit, reducing it by about 10\% for the single-plane barrier. Fifth,
we saw a dramatic deviation from the quasiclassical predictions as the
barrier becomes insulating due to strong electron correlations.  The 
characteristic voltage remains high for a larger range of Thouless energy
than in metallic junctions, and then decreases very rapidly as the barrier
passes through a critical thickness where the metal-insulator transition
occurs.  This behavior leads to the possibility of ``intrinsic pinholes.''

This work shows that correlation effects, and the interplay between 
superconductivity and oscillations brought on by the underlying Fermi
surface, play increasingly important roles in short-coherence-length
superconductors. In particular, optimizing the characteristic voltage of
a junction near the metal-insulator transition is possible, but requires
a careful tuning of the thickness of the barrier, the proximity to the
metal-insulator transition, and the operating temperature of the 
device.   Correlated insulating barriers can mimic effects due to pinholes
because the Josephson coupling depends very strongly on the thickness
leading to an ``intrinsic'' pinhole effect.
In the future, we plan on extending this work to 
$d$-wave superconductors for direct applications to high-$T_c$ superconductors.

\section{Acknowledgements}
We are grateful to the Office of Naval Research for funding under grant 
number N00014-99-1-0328. Real-axis analytic continuation calculations
were partially supported by the National Computational Science Alliance
under grant number DMR990007N (utilizing the NCSA SGI/CRAY ORIGIN 2000)
and were partially supported by a grant of HPC time from the Arctic
Region Supercomputer Center. We wish to acknowledge useful discussions 
with  T. Van Duzer, J. Ketterson, T. Klapwijk, J. Luine, J. Mannhart, 
I. Nevirkovets, N. Newman, J. Rowell, and S. Tolpygo. J.K.F. thanks the 
hospitality of the IBM, Almaden Research Center, where this work was 
completed.

\end{document}